\documentclass[12pt]{article}
\usepackage{amssymb}
\usepackage{amsmath}
\usepackage{epsfig}
\usepackage{graphicx}
\usepackage{wrapfig}

\begin{document}

\newcommand{\udbarxx}{$u \bar d \to \widetilde{\chi}_1^+ \widetilde{\chi}_2^0$ }
\newcommand{\ubardxx}{$\bar u  d \to \widetilde{\chi}_1^- \widetilde{\chi}_2^0$ }
\newcommand{\udbarxxr}{$u \bar d \to \widetilde{\chi}_1^+ \widetilde{\chi}_2^0 \gamma$ }
\newcommand{\udbarxxg}{$u \bar d \to \widetilde{\chi}_1^+ \widetilde{\chi}_2^0 g$ }
\newcommand{\EW}{${\cal{O}}(\alpha_{ew})~$}
\newcommand{\ppudbarxx}{$p \bar p \to u \bar d \to \widetilde{\chi}_1^+ \widetilde{\chi}_2^0$ }
\newcommand{\ppxx}{$p p \to \widetilde{\chi}_1^{\pm} \widetilde{\chi}_2^0$ }
\newcommand{\ppbarxx}{$p \bar p \to \widetilde{\chi}_1^{\pm} \widetilde{\chi}_2^0$ }
\newcommand{\ppppbarudbarxx}{$p \bar p/pp \to \widetilde{\chi}_1^\pm \widetilde{\chi}_2^0$ }

\title{Full one-loop electroweak and NLO QCD
corrections to the associated production of chargino and
neutralino at hadron colliders \footnote{Supported by National
Natural Science Foundation of China.}} \vspace{3mm}

\author{\small{ Sun Hao$^{2}$, Han Liang$^{2}$, Ma Wen-Gan$^{1,2}$,
Zhang Ren-You$^{2}$, Jiang Yi$^{2}$, and Guo Lei$^{2}$ }\\
{\small $^{1}$ CCAST (World Laboratory), P.O.Box 8730, Beijing 100080, P.R.China} \\
{\small $^{2}$ Department of Modern Physics, University of Science and Technology}\\
{\small of China (USTC), Hefei, Anhui 230027, P.R.China}  }

\date{}
\maketitle \vskip 12mm

\begin{abstract}

We study the process of the association production of chargino and
neutralino including the NLO QCD and the complete one-loop
electroweak corrections in the framework of the minimal
supersymmetric standard model(MSSM) at the Fermilab Tevatron and
the CERN Large Hadron Collider (LHC). In both the NLO QCD and
one-loop electroweak calculations we apply the algorithm of the
phase-space slicing(PSS) method. We find that the NLO QCD
corrections generally increase the Born cross sections, while the
electroweak relative corrections decrease the Born cross section
in most of the chosen parameter space. The NLO QCD and electroweak
relative corrections typically have the values of about 32$\%$ and
$-8\%$ at the Tevatron, and about 42$\%$ and $-6\%$ at the LHC
respectively. The results show that both the NLO QCD and the
complete one-loop electroweak corrections to the processes $p \bar
p/pp \to \widetilde{\chi}_1^{\pm} \widetilde{\chi}_2^0+X$ are
generally significant and should be taken into consideration in
precision experimental analysis.
\end{abstract}

\vskip 3cm

{\large\bf PASC:12.60.Jv, 14.80.Ly, 12.15.Lk, 12.38.Bx }

\vfill \eject

\baselineskip=0.36in

\renewcommand{\theequation}{\arabic{section}.\arabic{equation}}
\renewcommand{\thesection}{\Roman{section}}
\newcommand{\nb}{\nonumber}

\makeatletter      
\@addtoreset{equation}{section}
\makeatother       

\section{Introduction}
\par
People believe that the minimal supersymmetric standard
model(MSSM) is a very attractive extension of the standard
model(SM). The direct discovery of the supersymmetric particles is
one of the most important endeavors of present and future high
energy experiments. The MSSM theory predicts many new particles,
such as: squarks, sleptons, neutral CP-even Higgs-bosons $h^0$,
$H^0$ and CP-odd Higgs boson $A^0$, charged Higgs-bosons
$H^{\pm}$, charginos $\widetilde{\chi}^\pm_i(i=1,2)$, which are
the mixtures of charged winos and charged higgsinos, and four
neutralinos $\widetilde{\chi}^0_j(j=1-4)$ being the mixtures of
the neutral wino, bino and two neutral higgsinos.

\par
In most of the constrained MSSM scenarios, such as the
R-conserving minimal supergravity(mSUGRA), the lighter
chargino($\widetilde \chi^\pm _1$) and the
neutralinos($\widetilde\chi^0_1,\widetilde\chi^0_2$) are
considerably less massive than the gluinos and squarks over most
of the parameter space, they may belong to the class of the
relative light supersymmetric particles. The association
production of $\widetilde\chi^\pm_1$ and $\widetilde\chi^0_2$,
which seems to be one of the primary source of the trilepton
events, is a promising channel for supersymmetric particle
searches at hadron colliders. In the association production of the
lighter chargino($\widetilde \chi^\pm _1$) and the second lightest
nuetralino($\widetilde\chi^0_2$), both final supersymmetric
particles could dominantly decay to leptons, i.e.,
$\widetilde{\chi}^{\pm}_1 \to \widetilde \chi ^0_1 \cal
\ell^{\pm}\nu_\ell $ and $\widetilde{\chi}^0_2 \to \widetilde \chi
^0_1 \cal \ell^{+}\ell^-$, which leads to a gold-plated
$\ell^{\pm}\ell^{+}\ell^{-}$ trilepton
signature\cite{trilepton0}\cite{trilepton1}. We expect that the
lighter charginos and neutralinos to be detected in their above
mentioned decays at the Tevatron and the LHC. Actually, such
trilepton signature was exploited primarily in the CDF and D0
experiments at the Tevatron and got the similar bounds to those
from LEP2 in the MSSM parameter space\cite{trilepton2}.

\par
To investigate the discovering potential of hadron colliders, not
only a proper understanding of the hadron production mechanisms is
necessary, but also accuracy theoretical predictions of the
signature should be provided. As we know, the impact of higher
order electroweak and QCD contributions normally grows with
increasing colliding energy and would become to be more obvious at
very high energies. In Ref.\cite{Barger}, V. Barger and Chung Kao
investigated the prospects for detecting trilepton events
$(l=e~or~\mu)$ from neutralino-chargino$(\tilde\chi^0_2
\tilde\chi^{\pm}_1)$ associated production at the upgraded
Tevatron in the mSUGRA model. They found that if there is a large
integrated luminosity(for example ${\cal L}=30~fb^{-1}$) and the
decay modes $\widetilde{\chi}^{\pm}_1 \to \widetilde \chi ^0_1
\cal \ell^{\pm}\nu_\ell $ and $\widetilde{\chi}^0_2 \to \widetilde
\chi ^0_1 \cal \ell^{+}\ell^-$ are kinematically dominant, the
value of statistical significance ($N_S \equiv S/\sqrt{B}$,
$S=$number of signal events, and $B=$number of background events)
can reach $36.9$ when we use suitable cuts and take $\tan\beta=3$.
Then one can expect the accuracy of the cross section measurement
of the neutralino-chargino$(\tilde\chi^0_2 \tilde\chi^{\pm}_1)$
associated production at the upgraded Tevatron can reach few
percent. Therefore, for the precise experiments at $TeV$ scale
hadron colliders, both the higher order QCD and electroweak
corrections should be considered in the theoretical predictions,
and thereby one can improve experimental mass bounds and exclusion
limits for the new particles. Moreover, the consideration of
higher order QCD contributions can reduce the dependence of the
cross sections on the renormalization and factorization scales in
the LO. And the cross sections in NLO are under much better
theoretical control than the leading order estimates.

\par
There have been many works which present the theoretical
calculations of the production of SUSY particles in hadron
collisions at NLO QCD level, such as,
Refs.\cite{SUSYQCD1}\cite{SUSYQCD2}\cite{SUSYQCD3}\cite{SUSYQCD4}.
In Ref.\cite{SUSYQCD4} it presents the complete next-to-leading
order SUSY QCD analysis for the production of all possible pairs
of noncolored supersymmetric particles, including $pp/p\bar p \to
\widetilde{\chi}^{\pm}_1 \widetilde{\chi}^{0}_2+X $. In their
calculations of the NLO QCD corrections, the infrared and
collinear singularities were extracted by applying the dipole
subtraction method\cite{Catani}.

\par
In our work, we are to calculate and discuss the complete one-loop
electroweak radiative corrections to the processes $p\bar p/ pp
\to \widetilde{\chi}^{\pm}_1 \widetilde{\chi}^{0}_2+X $ at the
Tevatron and the LHC. And for the completeness of our
investigation we present also the NLO QCD corrections to the
$p\bar p/ pp \to \widetilde{\chi}^{\pm}_1 \widetilde{\chi}^{0}_2+X
$ processes by applying the phase-space slicing
method(PSS)\cite{PSS}. At the same time we compare our NLO QCD
numerical results from the PSS method with those from dipole
substraction method in Ref.\cite{SUSYQCD4}.

\par
The structure of this paper is organized as follows: In Section 2,
we calculate the cross section of the leading order results for
the subprocess \udbarxx. In Section 3, we give the analytical and
numerical calculations and discussions for the NLO QCD corrections
to the processes $p\bar p/ pp \to \widetilde{\chi}^{\pm}_1
\widetilde{\chi}^{0}_2+X $ at the Tevatron and the LHC. In Section
4, we present the calculations of the complete one-loop
electroweak corrections and discussions for the process $p\bar p/
pp \to \widetilde{\chi}^{\pm}_1 \widetilde{\chi}^{0}_2+X $.
Finally, a short summary is given.

\vskip 5mm
\section{The leading order calculation for subprocess \udbarxx}
\par
\begin{figure}[hbtp]
\vspace*{-1cm} \centerline{ \epsfxsize = 8cm \epsfysize = 3cm
\epsfbox{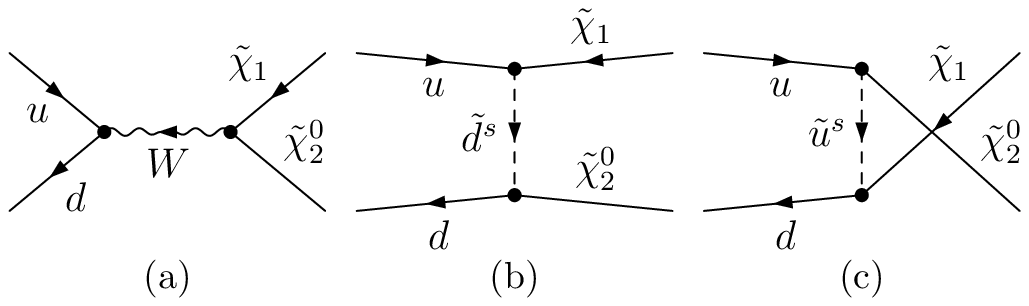}}  \vspace*{0cm}\caption{\em The tree-level
Feynman diagrams for the subprocess \udbarxx.}
\label{udbarxx_tree}
\end{figure}

Since the cross sections for the subprocess \udbarxx and its
charge-conjugate subprocess \ubardxx in the CP-conserved MSSM are
the same, we present here only the calculation of the subprocess
\udbarxx. The tree-level diagrams for the subprocess \udbarxx are
shown in Fig.\ref{udbarxx_tree}.

\vspace*{1.5cm}

\par
In our calculation, we neglect the small masses of the
light-quarks and there is no contributions from the Feynman
diagrams which involve the couplings $H^{+}/G^{+}-u-d$.

\par
We denote the subprocess as
\begin{equation}
u(p_1)+\bar d (p_2) \to \widetilde{\chi}_1^+ (k_3) +
\widetilde{\chi}_2^0(k_4)
\end{equation}
where $p_i~(i = 1, 2)$ and $k_i~(i = 3, 4)$ are the four-momenta
of incoming up-quark/anti-down-quark and outgoing
$\widetilde{\chi}_1^+$ and $\widetilde{\chi}_2^0$, respectively.
All these four-momenta satisfy the on-shell conditions: $k_3^2 =
m_{\widetilde{\chi}_1^+}^2$, $k_4^2 = m_{\widetilde{\chi}_2^0}^2$,
$p_1^2 = p_2^2 = 0$. The center-of-mass energy squared is denoted
by $\hat{s} = E_{cm}^2 = (p_1+p_2)^2$.

\par
The amplitude of \udbarxx subprocess can be divided into three
parts and expressed as
\begin{eqnarray}
{\cal M}^{0}&=&{\cal M}^{s}+{\cal M}^{t}+{\cal M}^{u}
\end{eqnarray}
where ${\cal M}^{0}$ is the tree-level amplitude and ${\cal
M}^{s}$, ${\cal M}^{t}$, ${\cal M}^{u}$ represent the amplitude
parts arising from the s-channel, t-channel and u-channel diagrams
shown in Fig.\ref{udbarxx_tree}(a-c), respectively. Then the
lowest order cross section for the subprocess \udbarxx in the MSSM
is obtained by using the following formula:
\begin{eqnarray}
\hat{\sigma}_{0}(\hat{s},u \bar d \to \widetilde{\chi}_1^+
\widetilde{\chi}_2^0) =
\frac{1}{16\pi\hat{s}^2}\int^{\hat{t}_{max}}_{\hat{t}_{min}}d\hat{t}\overline\sum|{\cal
M}^0|^2.\label{summation}
\end{eqnarray}
where
\begin{eqnarray}
\label{tmaxmin} \hat{t}_{min}(\hat{t}_{max})=\frac{1}{2}\left \{(
m^2_{\widetilde\chi^+_1} + m^2_{\widetilde\chi^0_2} - \hat{s} )
 \pm  \sqrt{ (\hat{s}-m^2_{\widetilde\chi^+_1} -
 m^2_{\widetilde\chi^0_2} )^2 - 4m^2_{\widetilde\chi^+_1} m^2_{\widetilde\chi^0_2}
 }\right\}.
\end{eqnarray}
The summation is taken over the spins and colors of initial and
final states, and the bar over the summation denotes averaging
over the spins and colors of initial partons.

\par
\section{NLO QCD corrections}
\par
In considering the NLO QCD corrections to the subprocesses $u \bar
d(\bar u d) \to \widetilde{\chi}_1^{\pm} \widetilde{\chi}_2^0$, we
should involve the gluon emission subprocess \udbarxx$+g$ to
cancel the soft IR divergence arising from the virtual QCD
corrections of the subprocesses $u \bar d(\bar u d) \to
\widetilde{\chi}_1^{\pm} \widetilde{\chi}_2^0$. The light-quark
emission subprocesses $g u (\bar d) \to \widetilde{\chi}_1^+
\widetilde{\chi}_2^0+d (\bar u)$ should be also included for a
consistent and complete mass factorization.

\vskip 5mm
\subsection{Virtual corrections}

\par
\begin{figure}[hbtp]
\vspace*{-1cm} \centerline{ \epsfxsize = 15cm \epsfysize = 12cm
\epsfbox{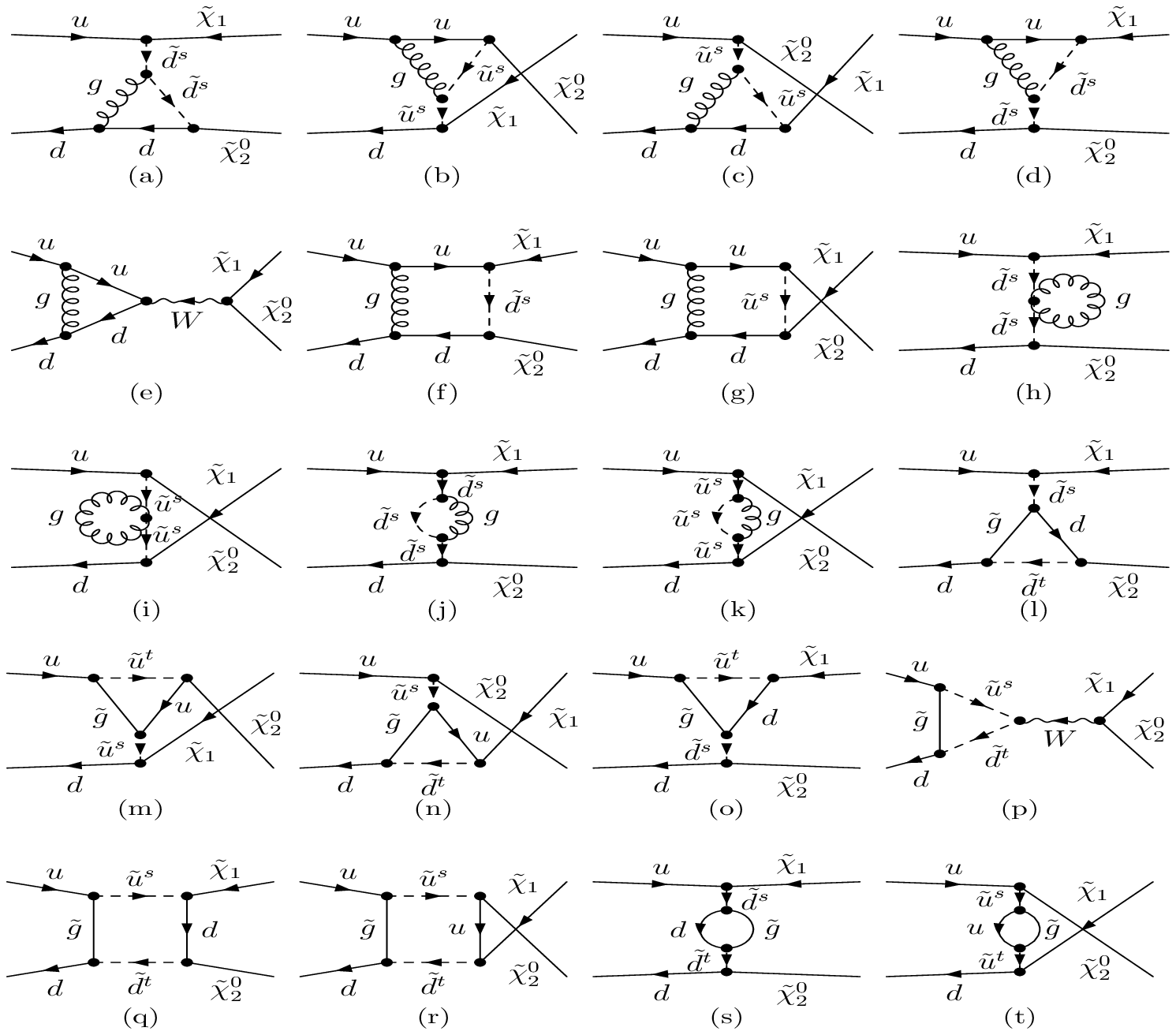}}  \vspace*{-0.5cm}\caption{The one-loop Feynman
diagrams of virtual QCD corrections for the subprocess \udbarxx in
the MSSM.} \label{virtual_QCD}
\end{figure}

The complete one-loop Feynman diagrams of QCD corrections in the
MSSM, which are built up with gluon, gluino, quark and squark
exchanging loops, are shown in Fig.\ref{virtual_QCD}. We use the
fermion flow prescription\cite{Majorana} for the calculation of
the matrix elements including Majorana particles. Same as in the
tree-level calculation, we neglect the masses of the light-quarks
in the calculation of the virtual QCD corrections for the
subprocess \udbarxx. There exist both ultraviolet(UV) divergences
and soft/collinear IR singularities in the amplitudes from these
QCD one-loop diagrams. We use the method of the phase-space
slicing(PSS)\cite{PSS} to treat the soft and collinear
divergences. The PSS method is intuitive, simple to implement, and
relies on a minimum of process dependent information and has been
used in many works. In the NLO QCD calculation, we adopt the 't
Hooft-Feynman gauge and dimensional regularization method with
$n=4-2\epsilon$ to evaluate the one-loop contributions. The
expressions for the relevant renormalization constants can be
found in next section, except all the relevant self-energies
including only the QCD parts instead of the electroweak parts. We
have verified the cancellation of the UV divergence in the virtual
QCD corrections analytically. Then we get an UV finite amplitude
for the ${\cal O}(\alpha_{s})$ virtual radiative corrections.

\vspace*{1.0cm}

\par
The virtual correction to the cross section can be written as
\begin{eqnarray}
\hat \sigma^{V}(\hat{s},u \bar d \to \widetilde{\chi}_1^+
\widetilde{\chi}_2^0) =
\frac{1}{16\pi\hat{s}^2}\int^{\hat{t}_{max}}_{\hat{t}_{min}}d\hat{t}~~2Re\overline\sum[({\cal
M}^V)^\dagger {\cal M}^0].
\end{eqnarray}
where ${\cal M}^V$ is the UV renormalized amplitude for virtual
corrections, and again the summation with bar over head means the
same operation as before. The expressions for $\hat{t}_{max}$ and
$\hat{t}_{ min}$ can be found in Eq.(\ref{tmaxmin})

\par
After the renormalization procedure, $\hat\sigma^V$ is UV-finite.
However, it still contains the soft/collinear IR singularities.
The IR divergence part in $d\hat \sigma^{V}$ can be obtained as
\begin{eqnarray}
\label{virtual cross section} d\hat \sigma^{V}|_{IR} =
d\hat\sigma^0 \left
[\frac{\alpha_s}{2\pi}\frac{\Gamma(1-\epsilon)}{\Gamma(1-2\epsilon)}
       \left (\frac{4\pi \mu_r^2}{\hat s}\right )^\epsilon \right ]
       \left (\frac{A^V_2}{\epsilon^2}+\frac{A^V_1}{\epsilon}\right ),
\end{eqnarray}
where
\begin{eqnarray}
A^V_2 = -\frac{8}{3},~~~~A^V_1 = -4.
\end{eqnarray}

\par
The soft divergences can be cancelled by adding the soft real
gluon emission corrections. The collinear divergences together
with those coming from the real light-quark emission corrections
are absorbed into the parton distribution functions, which will be
discussed in the next subsection.

\vskip 5mm
\subsection{Real emission corrections}
\par
The real emission subprocesses, which present NLO QCD corrections
to the \udbarxx subprocess, include real gluon emission subprocess
\udbarxx$+g$, and real light-quark emission subprocesses $g u \to
\widetilde{\chi}_1^+ \widetilde{\chi}_2^0+d$, $g \bar d \to
\widetilde{\chi}_1^+ \widetilde{\chi}_2^0+\bar u$. The later two
subprocesses will make additional contributions to the final
two-body and the three-body cross sections.

\par
{\bf 1. Real gluon emission corrections}
\par
The real gluon emission subprocess \udbarxx$+g$ presents ${\cal
O}(\alpha_s)$ corrections to the subprocess \udbarxx. It is also
one of the origins of IR singularities. Its IR singularities can
be either of soft or collinear nature and can be conveniently
isolated by slicing the \udbarxx$+g$ phase space into different
regions defined by suitable cutoffs, a method which goes under the
general name of phase-space slicing(PSS)\cite{PSS}. The soft IR
singularity part from subprocess \udbarxx$+g$ cancels exactly the
analogous singularity presented in the ${\cal O}(\alpha_s)$
virtual corrections calculated in above subsection.

\vspace*{1.5cm}
\begin{figure}[hbtp]
\vspace*{-1cm} \centerline{ \epsfxsize = 11cm \epsfysize = 5cm
\epsfbox{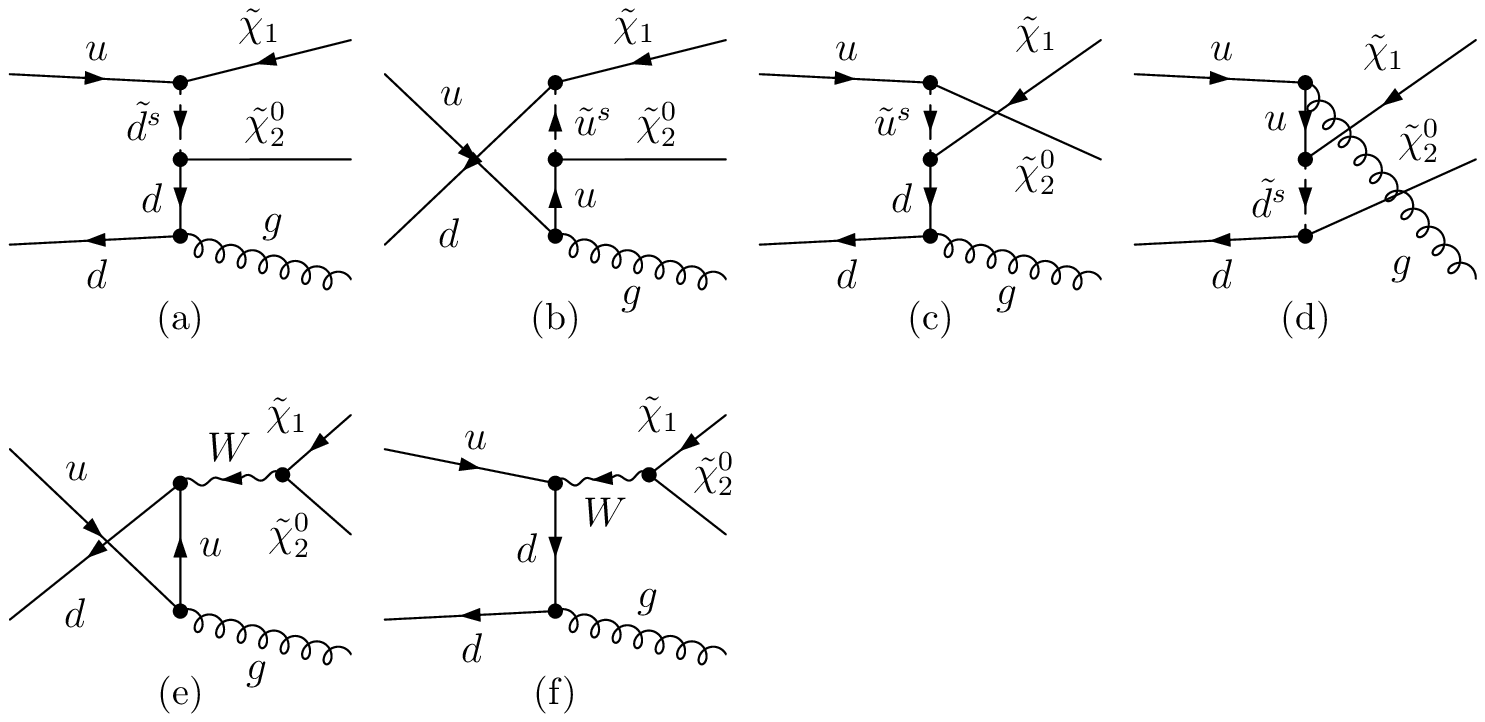}}  \vspace*{-0.5cm}\caption{The tree-level
Feynman diagrams for the real gluon emission subprocess
\udbarxx+g.} \label{udbarxxg}
\end{figure}

\par
We denote the $2 \to 3$ subprocess \udbarxx$+g$ as
\begin{equation}
u(p_1)+\bar d (p_2) \to \widetilde{\chi}_1^+ (k_3) +
\widetilde{\chi}_2^0(k_4) + g(k_5),
\end{equation}
The Mandelstam variables are defined as
\begin{eqnarray}
\hat s&=& (p_1+p_2)^2, ~~\hat t = (p_1-k_3)^2, ~~\hat u = (p_1-k_4)^2,~~\hat s_{45}=(k_4+k_5)^2,   \nb\\
 \hat t_{15}&=&(p_1-k_5)^2,~~ \hat
t_{25}=(p_2-k_5)^2,~~\hat t_{45}=(k_4-k_5)^2
\end{eqnarray}
By introducing an arbitrary small soft cutoff $\delta_s$ we
separate the phase space of the $2 \to 3$ subprocess \udbarxx$+g$
into two regions, according to whether the energy of the emitted
gluon is soft, i.e. $E_5\leq \delta_s \sqrt{\hat s}/2$, or hard,
i.e. $E_5
> \delta_{s}\sqrt{\hat s}/2 $. The partonic cross section can be
written as
\begin{eqnarray}
\hat \sigma^R_g (u \bar d \to \widetilde{\chi}_1^+
\widetilde{\chi}_2^0+g)=\hat \sigma^S_g(u \bar d \to
\widetilde{\chi}_1^+ \widetilde{\chi}_2^0+g)+\hat \sigma^H_g (u
\bar d \to \widetilde{\chi}_1^+ \widetilde{\chi}_2^0+g),
\end{eqnarray}
where $\hat \sigma ^S_g$ is obtained by integrating over the soft
gluon emission phase space region, and contains all the soft IR
singularities. In order to isolate the remaining collinear
singularities from $\hat\sigma ^H_g$, we further decompose
$\hat\sigma^H_g$ into hard-collinear(HC) and hard non-collinear
($\overline{HC}$) parts by introducing another cutoff $\delta_c$
named collinear cutoff,
\begin{eqnarray}
\hat \sigma^{H}_g (u \bar d \to \widetilde{\chi}_1^+
\widetilde{\chi}_2^0+g)=\hat \sigma^{HC}_g(u \bar d \to
\widetilde{\chi}_1^+ \widetilde{\chi}_2^0+g)+\hat
\sigma^{\overline{HC}}(u \bar d \to \widetilde{\chi}_1^+
\widetilde{\chi}_2^0+g).
\end{eqnarray}
The HC regions of the phase space are those in collinear
condition, where any invariant, $\hat{t}_{15}, \hat{t}_{25},
\hat{t}_{45}$, becomes smaller in magnitude than $\delta_c
\sqrt{\hat s}$, while at the same time the emitted gluon remains
hard. $\hat\sigma^{HC}_g$ contains the collinear divergences. In
the soft and HC regions, $\hat\sigma^S_g$ and $\hat\sigma^{HC}_g$
can be obtained by performing the phase space integration in
$n=4-2\epsilon$ dimensions analytically. But in the
$\overline{HC}$ region, $\hat\sigma^{\overline{HC}}_g$ is finite
and can be evaluated in four dimensions by using standard Monte
Carlo techniques\cite{MC}. The cross sections, $\hat\sigma^{S}_g$,
$\hat\sigma^{HC}_g$ and $\hat\sigma^{\overline{HC}}_g$, depend on
the arbitrary parameters, $\delta_s$ and $\delta_c$.

\par
With the arbitrary small cutoff $\delta_s$, the differential cross
section in the soft region is given as
\begin{equation}\label{soft cross section1}
d\hat\sigma^S_g = d\hat \sigma^0 \left [\frac{\alpha_s}{2\pi}
            \frac{\Gamma(1-\epsilon)}{\Gamma(1-2\epsilon)}
        \left(\frac{4\pi \mu_r^2}{\hat s}\right )^\epsilon \right ]
        \left (\frac{A^S_2}{\epsilon^2}+\frac{A^S_1}{\epsilon}+A^S_0\right ),
\end{equation}
with
\begin{eqnarray}
A^S_2 = \frac{8}{3},~~ A^S_1 = -\frac{16}{3}\ln \delta_s,~~ A^S_0
= \frac{16}{3}\ln^2 \delta_s.
\end{eqnarray}
\par
In the limit where two of the partons are collinear, the three
body phase space is greatly simplified. And at the same limit, the
leading pole approximation of the matrix element is valid.
According to whether the collinear singularities are initial or
final state in origin, where $\hat \sigma^{HC}_g$ can be separated
into two pieces
\begin{eqnarray}
\hat \sigma^{HC}_g=\hat \sigma^{HC}_{g,i}+\hat \sigma^{HC}_{g,f}
\end{eqnarray}
Since in the process \udbarxx +g, only initial particles are
involved in strong interaction and massless in the limit, we only
need to calculate the cross section $\hat \sigma^{HC}_{g,i}$ which
arises from the case that the emitted gluon is collinear to the
initial partons, i.e., $0\leq t_{15},t_{25}\leq \delta_c \hat s$.

\par
In $HC$ phase space region, the initial state partons $i(i=u, \bar
d)$ is considered to split into a hard parton $i^\prime$ and a
collinear gluon, $i \rightarrow i^\prime g$, with
$p_{i^\prime}=zp_{i}$ and $k_5=(1-z)p_{i}$. The cross section
$\sigma^{HC}_{g,i}$ can be written as
\begin{eqnarray}\nb
d \sigma^{HC}_g&=&d \sigma^{HC}_{g,i}=d\hat
\sigma^{0}\left[\frac{\alpha_s}{2\pi}
\frac{\Gamma(1-\epsilon)}{\Gamma(1-2\epsilon)}\left(\frac{4\pi\mu^2_r}{\hat
s}\right)^\epsilon\right]\left(-\frac{1}{\epsilon}\right)\delta_c^{-\epsilon}
\left[P_{uu}(z,\epsilon)G_{u/A}(x_A/z)G_{\bar
d /B}(x_B)+\right.\\
&& \left. P_{\bar d\bar d}(z,\epsilon)G_{\bar
d/A}(x_A/z)G_{u/B}(x_B)+(A\leftrightarrow B)\right
]\frac{dz}{z}\left(\frac{1-z}{z}\right)^{-\epsilon}dx_Adx_B
\label{HCi}
\end{eqnarray}
where $G_{u,\bar d/A,B}$ are the bare parton distribution
functions. A and B refer to protons at the LHC, and proton,
antiproton at the Tevatron. $P_{uu}(z,\epsilon)$ and $P_{\bar
d\bar d}(z,\epsilon)$ are the n-dimensional unregulated ($z <1$)
splitting functions related to the usual Altarelli-Parisi
splitting kernels\cite{APsk}. $P_{ii}(z,\epsilon) (i=u, \bar d)
$can be written explicitly as
\begin{eqnarray}\nb
\label{Peq0} P_{ii}(z,\epsilon)&=& P_{ii}(z) + \epsilon
P'_{ii}(z)\\\nb
P_{ii}(z) &=& C_F \frac{1+z^2}{1-z}\\
P'_{ii}(z) &=& -C_F(1-z)~~~~~(i=u, \bar d),   \label{Pii}
\end{eqnarray}
with $C_F$=4/3.

\par
{\bf 2. Real light-quark emission corrections of $g u (\bar d) \to
\widetilde{\chi}_1^+ \widetilde{\chi}_2^0+d (\bar u)$}

\par
\begin{figure}[hbtp]
\vspace*{-1cm} \centerline{ \epsfxsize = 11cm \epsfysize = 5cm
\epsfbox{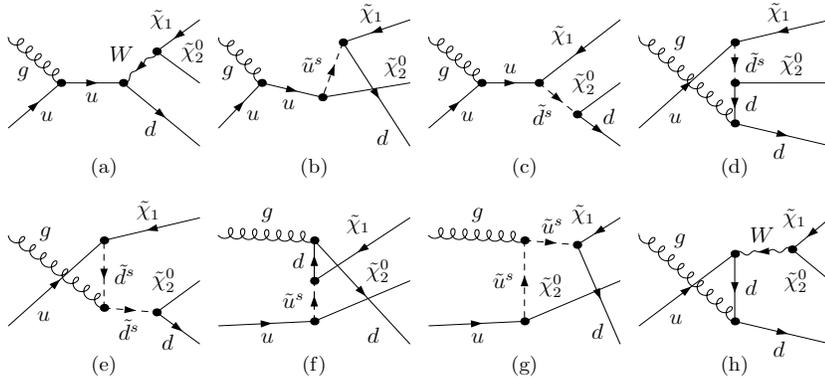}}  \vspace*{-0.5cm}\caption{The tree-level
Feynman diagrams for the real light-quark emission subprocess $g u
\to \widetilde{\chi}_1^+ \widetilde{\chi}_2^0+d $.}
\label{Feyguxxd}
\end{figure}
\begin{figure}[hbtp]
\vspace*{-1cm} \centerline{ \epsfxsize = 11cm \epsfysize = 5cm
\epsfbox{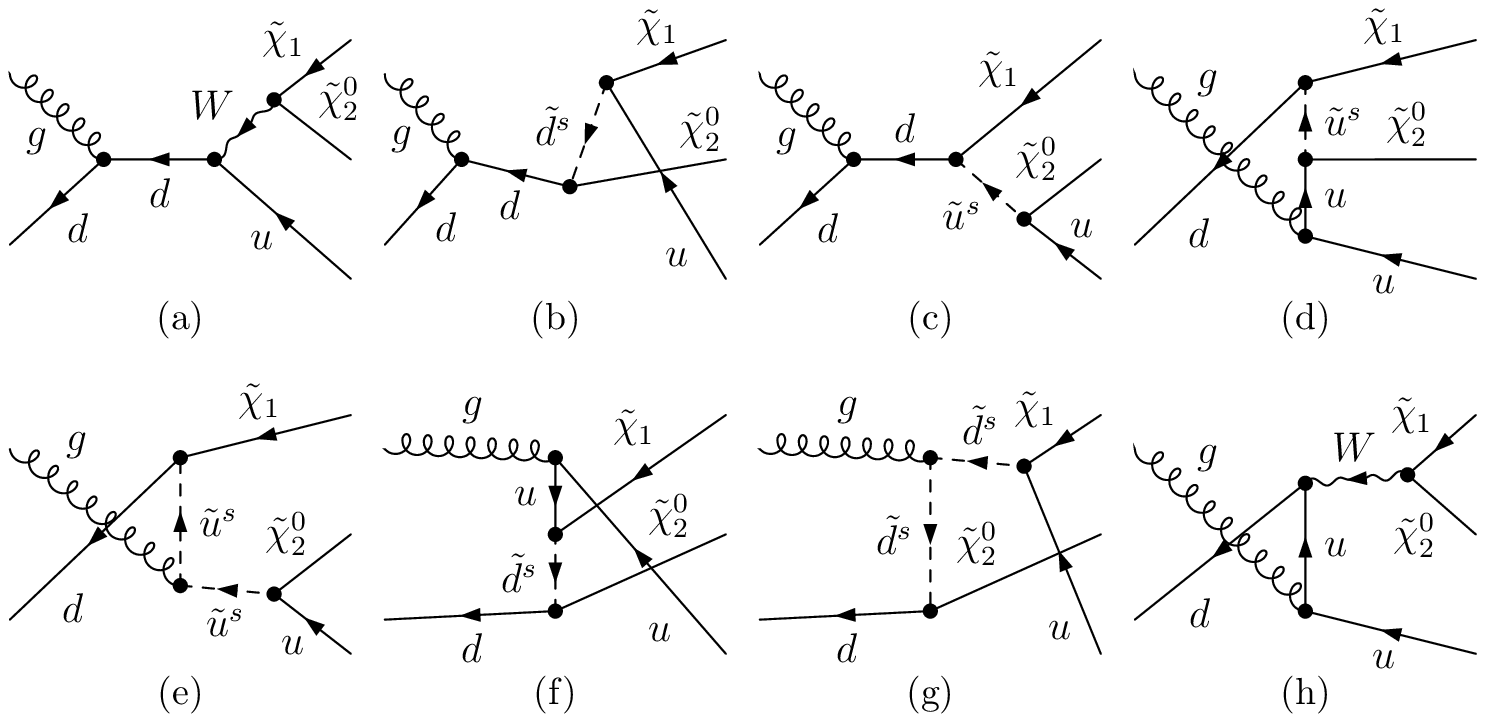}}  \vspace*{-0.5cm}\caption{The tree-level
Feynman diagrams for the real light-quark emission subprocess $g
\bar d \to \widetilde{\chi}_1^+ \widetilde{\chi}_2^0+\bar u$.}
\label{Feygdbarxxubar}
\end{figure}

In addition to the real gluon emission subprocess, the
subprocesses $g u (\bar d) \to \widetilde{\chi}_1^+
\widetilde{\chi}_2^0+d (\bar u)$ should also be included. The
contributions from these two subprocesses contain only the initial
state collinear singularity. These subprocesses will make
additional contributions to the two-body term cross section
$\sigma^{(2)}$ and the three-body term cross section
$\sigma^{(3)}$. The Feynman diagrams for these two subprocesses at
the tree-level are shown in Fig.\ref{Feyguxxd} and
Fig.\ref{Feygdbarxxubar}, respectively.

\vspace*{1.5cm}

\vspace*{1.5cm}

\par
By using the PSS method described above, we split the phase space
into two regions: collinear region and non-collinear region.
\begin{eqnarray}
\hat{\sigma}^R_{q}(g u (\bar d) \to \widetilde{\chi}_1^+
\widetilde{\chi}_2^0+d (\bar u)) &=& \hat{\sigma}^{\rm HC}_{q}(g u
(\bar
d) \to \widetilde{\chi}_1^+ \widetilde{\chi}_2^0+d (\bar u)) \nb \\
&+&\hat{\sigma}^{\overline{HC}}_{q}(g u (\bar d) \to
\widetilde{\chi}_1^+ \widetilde{\chi}_2^0+d (\bar u)).
\end{eqnarray}
Also $\hat{\sigma}^{\overline{HC}}_{q}$ in hard non-collinear
region is finite and can be evaluated in four dimensions using
Monte Carlo method. The differential cross sections of
$d\sigma^{HC}_{q}$ for subprocesses $g u (\bar d) \to
\widetilde{\chi}_1^+ \widetilde{\chi}_2^0+d (\bar u)$ can be
written as
\begin{eqnarray}
\label{collinear-b} &&d\sigma^{HC}_q (g u (\bar d) \to
\widetilde{\chi}_1^+ \widetilde{\chi}_2^0+d (\bar u))= \nb \\
&&d\hat{\sigma}^0 \left[\frac{\alpha_s}{2 \pi}
\frac{\Gamma(1-\epsilon)}{\Gamma(1-2 \epsilon)}\left(\frac{4 \pi
\mu_r^2}{\hat{s}}\right)^{\epsilon}\right] \left
(-\frac{1}{\epsilon}\right )\delta_c^{-\epsilon} [
P_{d(\bar u)g}(z,\epsilon)G_{g/A}(x_A/z)G_{u(\bar d)/B}(x_B) \nb \\
&&+(A\leftrightarrow B)]\frac{dz}{z}\left( \frac{1-z}{z}\right
)^{-\epsilon }dx_Adx_B. ~~~~~~~~~~~~~~~~~~~~~~~~~~~~~~~~~~~~~~
\end{eqnarray}
with $P_{ij}(z,\epsilon) (i=\bar u,d,~j=g) $ expressed explicitly
as

\begin{eqnarray}
\label{Peq1}
P_{d(\bar u)g}(z,\epsilon)&=&P_{d(\bar u)g}(z)+ \epsilon P'_{d(\bar u)g}(z), \nb \\
P_{d(\bar u)g}(z)&=&\frac{1}{2}[z^2+(1-z)^2],~~~~~~ P'_{d(\bar
u)g}(z) = -z(1-z).
\end{eqnarray}

\par
{\bf 3. NLO QCD corrected cross section for $pp /p\bar p \to
\widetilde{\chi}_1^\pm \widetilde{\chi}_2^0+X$}

\par
After adding the renormalized virtual corrections and the real
corrections, the partonic cross sections still contain the
collinear divergences, which can be absorbed into the redefinition
of the distribution functions at NLO. Using the $\overline{\rm
MS}$ scheme, the scale dependent NLO parton distribution functions
are given as \cite{Harris}
\begin{eqnarray}
G_{i/A}(x,\mu_f)=G_{i/A}(x)+\left (-\frac{1}{\epsilon}\right
)\left[\frac{\alpha_s}{2 \pi} \frac{\Gamma(1-\epsilon)}{\Gamma(1-2
\epsilon)}\left(\frac{4 \pi
\mu_r^2}{\mu_f^2}\right)^{\epsilon}\right]\int^1_z\frac{dz}{z}P_{ij}(z)G_{j/A}(x/z).
\nb \\
\end{eqnarray}
By using above definition, we get a NLO QCD parton distribution
function counter-terms which are combined with the hard collinear
contributions(Eq.(\ref{HCi}) and (\ref{collinear-b})) to result in
the $O(\alpha_s)$ expression for the remaining collinear
contributions:
\begin{eqnarray}
\label{collinear cross section1}
 d\sigma^{coll}&=&d\hat{\sigma}^0
\left[\frac{\alpha_s}{2 \pi} \frac{\Gamma(1-\epsilon)}{\Gamma(1-2
\epsilon)}\left(\frac{4 \pi
\mu_r^2}{\hat{s}}\right)^{\epsilon}\right] \{
\tilde{G}_{u/A}(x_A,\mu_f)G_{\bar{d}/B}(x_B,\mu_f)+G_{u/A}(x_A,\mu_f)\tilde{G}_{\bar{d}/B}(x_B,\mu_f)
\nb \\
&+& \sum_{\alpha=u,\bar{d}}\left [\frac{A_1^{sc}(\alpha \to \alpha
g)}{\epsilon}+A_0^{sc}(\alpha \to
\alpha g)\right ]G_{u/A}(x_A,\mu_f)G_{\bar{d}/B}(x_B,\mu_f) \nb \\
&+& (A\leftrightarrow B)\}dx_Adx_B,
\end{eqnarray}
where $A$ and $B$ are proton and antiproton for the Tevatron, and
proton and proton for the LHC, respectively.
\begin{eqnarray}
A_1^{sc}(u(\bar{d}) \to u(\bar{d}) g)&=& C_F(2 \ln \delta_s+3/2),
~~~A_0^{sc} = A_1^{sc} \ln\left (\frac{\hat{s}}{\mu_f^2}\right ),
\end{eqnarray}
and
\begin{eqnarray}
\tilde{G}_{\alpha/A,B}(x,\mu_f)=\sum_{c'=\alpha,g}\int^{1-\delta_s
\delta_{\alpha c'}}_x
\frac{dy}{y}G_{c'/A,B}(x/y,\mu_f)\tilde{P}_{\alpha
c'}(y),~~~(\alpha=u,\bar d)
\end{eqnarray}
with
\begin{eqnarray}
\tilde{P}_{\alpha c'}(y)=P_{\alpha c'} \ln\left
(\delta_c\frac{1-y}{y}\frac{\hat{s}}{\mu_f^2}\right )-P'_{\alpha
c'}(y).
\end{eqnarray}

\par
We can observe that the sum of the soft (Eq.(\ref{soft cross
section1})), collinear(Eq.(\ref{collinear cross section1})), and
ultraviolet renormalized virtual correction (Eq.(\ref{virtual
cross section})) terms is finite, i.e.,
\begin{eqnarray}
A^S_2&+&A^V_2=0, \nb \\
A^S_1&+&A^V_1+ A_1^{sc}(u\to ug)+A_1^{sc}(\bar d\to \bar dg)=0.
\end{eqnarray}
The final result for the total ${\cal O}(\alpha_s)$ correction
consists of two parts of contributions: a two-body term
$\sigma^{(2)}$ and a three-body term $\sigma^{(3)}$. The two-body
correction term $\sigma^{(2)}$ is expressed as
\begin{eqnarray}
\sigma^{(2)}&=&\frac{\alpha_s}{2 \pi} \int dx_Adx_Bd\hat{\sigma}^0
\{ G_{u/A}(x_A,\mu_f)G_{\bar{d}/B}(x_B,\mu_f)[A^S_0+A^V_0+A_0^{sc}(u\to ug)+A_0^{sc}(\bar d\to \bar dg)] \nb \\
&+& \tilde{G}_{u/A}(x_A,\mu_f)G_{\bar d/B}(x_B,\mu_f)
+G_{u/A}(x_A,\mu_f)\tilde{G}_{\bar d/B}(x_B,\mu_f)+(A
\leftrightarrow B ) \}.
\end{eqnarray}
And the three-body correction term $\sigma^{(3)}$ is written as
\begin{eqnarray}
\sigma^{(3)}&=&\sigma^{(3)}(pp/p\bar p \to u \bar d
\to \widetilde{\chi}_1^+ \widetilde{\chi}_2^0+g)  \nb \\
&+&\sigma^{(3)}(pp/p\bar p \to g u \to \widetilde{\chi}_1^+
\widetilde{\chi}_2^0+d)+\sigma^{(3)}(pp/p\bar p \to g \bar d \to
\widetilde{\chi}_1^+ \widetilde{\chi}_2^0+\bar u)
\nb \\
 &=&\int dx_Adx_B
[G_{u/A}(x_A,\mu_f)G_{\bar{d}/B}(x_B,\mu_f)+(A \leftrightarrow B
)]d\hat{\sigma}^{(3)}( u \bar d
\to \widetilde{\chi}_1^+ \widetilde{\chi}_2^0+g) \nb \\
&+&\int dx_Adx_B [G_{g/A}(x_A,\mu_f)G_{u/B}(x_B,\mu_f)+(A
\leftrightarrow B )]d\hat{\sigma}^{(3)}(g u
\to \widetilde{\chi}_1^+ \widetilde{\chi}_2^0+d) \nb \\
&+&\int dx_Adx_B [G_{g/A}(x_A,\mu_f)G_{\bar{d}/B}(x_B,\mu_f)+(A
\leftrightarrow B )]d\hat{\sigma}^{(3)} (g \bar d \to
\widetilde{\chi}_1^+ \widetilde{\chi}_2^0+\bar u).~~~~~~~~~`  \nb \\
\end{eqnarray}
Finally, the NLO total cross section for $pp/p\bar p \to
\widetilde{\chi}_1^+ \widetilde{\chi}_2^0+X$ is
\begin{eqnarray}
\sigma^{NLO}=\sigma^{0}+\sigma^{(2)}+\sigma^{(3)}.
\end{eqnarray}

\par
In the cross section part of $\hat\sigma^{(2)}+\hat\sigma^{(3)}$,
the dependence on the arbitrary cutoffs $\delta_s$ and $\delta_c$
should be vanished. This constitutes an important check of our
calculation.

\par
\subsection{Numerical results involving NLO QCD corrections}

\par
In Ref.\cite{SUSYQCD4}, W. Beenakker, {\rm et al,} presented the
NLO QCD calculations of the processes $pp/p\bar p \to
\widetilde{\chi}^{+}_1 \widetilde{\chi}^{0}_2+X $. There the
infrared and collinear singularities of the three parton cross
sections are extracted by applying algorithm of the dipole
subtraction method\cite{Catani}. As a check of our numerical
calculation by adopting the two cutoff PSS method, we take the
same mSUGRA input
parameters[$m_{1/2}=150~GeV,~m_0=100~GeV,~A_0=300~GeV,~\mu>0,~\tan\beta=4$]
and reproduce the Fig.2 shown in Ref.\cite{SUSYQCD4} with the
coincident numerical results. From this comparison we have
verified the correctness of our calculations of the NLO QCD
corrections. We take the colliding energies of $p \bar p/p p$ at
the Tevatron Run II and the LHC are 2 TeV and 14 TeV,
respectively.

\par
In the following numerical calculation for NLO QCD corrections, we
use the package FormCalc\cite{FeyArtsFormCalc} to get all the
masses of the supersymmetric particles by inputting $\tan\beta$,
$m_{A^0}$, $M_{susy}$, $\mu$, $M_2$ and $A_f$ parameters. In the
package FormCalc, the grand unification theory(GUT) relation $M_1
= (5/3)\tan^2 \theta_W M_2$ is adopted\cite{M1}, and
$M_{\tilde{Q}}=M_{\tilde{U}}=M_{\tilde{D}}=M_{\tilde{E}}=M_{\tilde{L}}=M_{susy}$
is assumed in the sfermion sector for simplification. We use the
one-loop formula for the running strong coupling constant
$\alpha_s$ with $\alpha_s(m_Z)=0.1187$, and take CTEQ6L and the
CTEQ6M parton distribution functions in calculating the LO and the
NLO cross sections, respectively\cite{CTEQ6}.

\par
\begin{figure}[hbtp] \centerline{
\epsfxsize = 6.8cm \epsfysize = 7cm \epsfbox{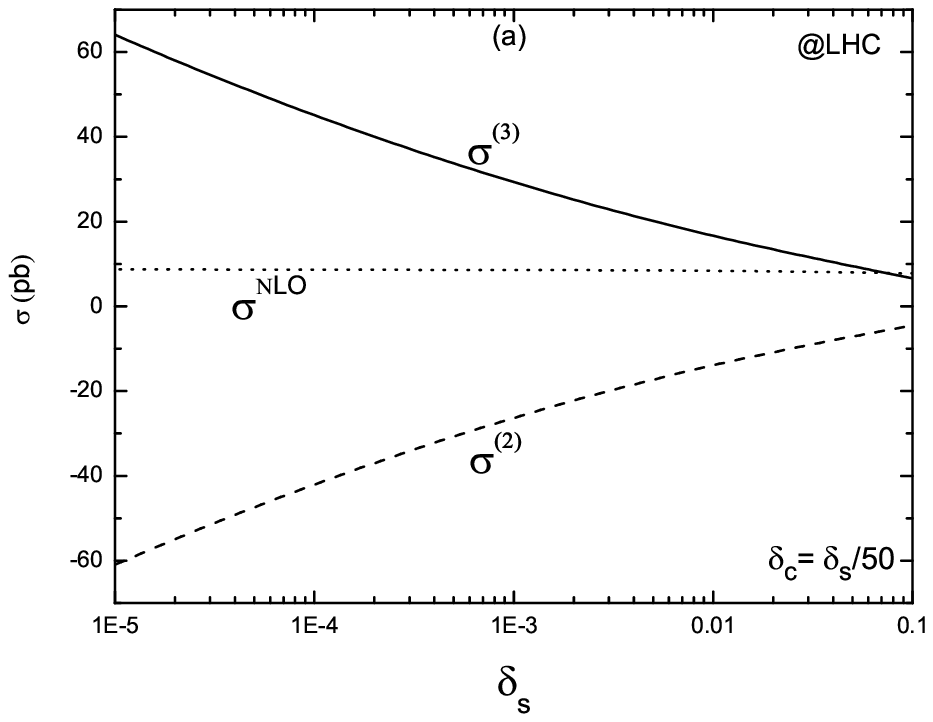} \epsfxsize
= 7.5cm \epsfysize = 7cm \epsfbox{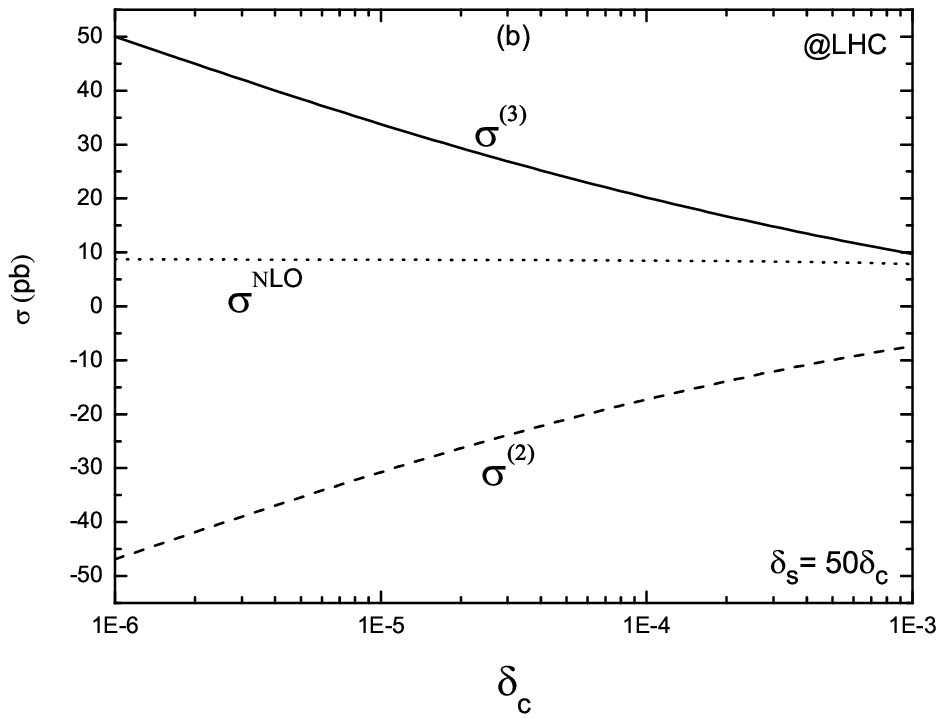}}\vspace*{0.2cm}
\centerline{ \epsfxsize = 6.8cm \epsfysize = 7cm
\epsfbox{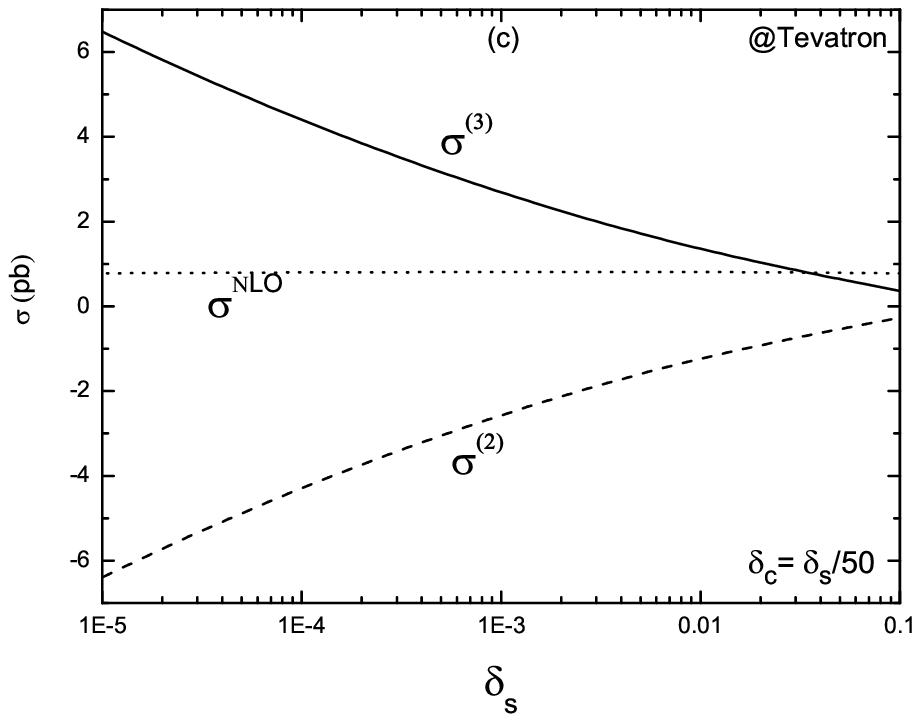} \epsfxsize = 7.5cm \epsfysize = 7cm
\epsfbox{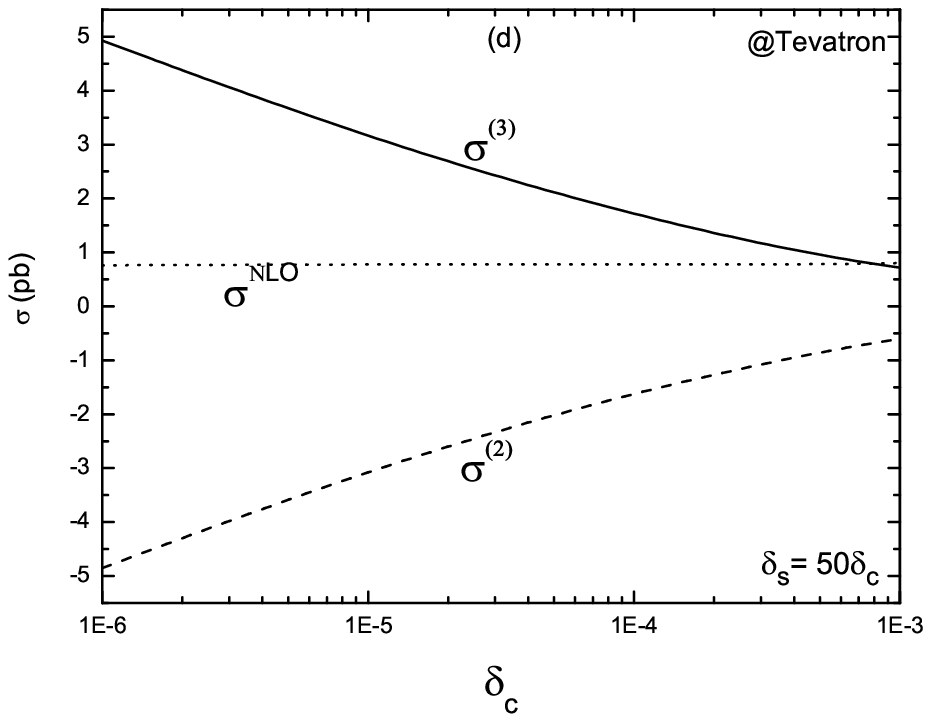}} \vspace*{-0.5cm}\caption{\em The dependence
of the total cross sections for $p \bar p/pp \to
\widetilde{\chi}_1^\pm \widetilde{\chi}_2^0+X$ processes at the
Tevatron and the LHC as the functions of the cutoff $\delta_s$
with $\delta_c=\delta_s/50$(see Fig.\ref{deltasc}(a),(c)) and
$\delta_c$ with $\delta_s=50\delta_c$(see
Fig.\ref{deltasc}(b),(d)), respectively.} \label{deltasc}
\end{figure}

Fig.\ref{deltasc} shows the independence of the NLO QCD corrected
cross sections on the arbitrary cutoffs $\delta_s$ and $\delta_c$
by applying the two cutoff PSS method. The two-body correction
term ($\sigma^{(2)}$) and three-body correction
term($\sigma^{(3)}$) and the NLO QCD corrected total cross section
$\sigma^{NLO}$ at the Tevatron and the LHC, are shown as the
functions of the soft cutoff $\delta_s$ with the collinear cutoff
$\delta_c=\delta_s/50$ (shown in Fig.\ref{deltasc}(a),(c)), and as
the functions of the collinear cutoff $\delta_c$ with the soft
cutoff $\delta_s=50\delta_c$ (shown in Fig.\ref{deltasc}(b),(d)).
The input supersymmetric parameters are taken as $\tan\beta=4$,
$m_{A^0}=300$ GeV, $M_{susy}=250$ GeV, $\mu=278$ GeV, $M_2=123$
GeV and $A_f=450$ GeV. We can see the NLO QCD corrected total
cross section $\sigma^{NLO}=\sigma^0+\delta
\sigma=\sigma^0+\sigma^{(2)}+\sigma^{(3)}$ is independent of the
cutoffs. This is an important check of the correctness of our
calculation. In the following numerical calculations, we set
$\delta_s=10^{-5}$ and $\delta_c=\delta_s/50$.

\vspace*{1.5cm}

\par
During our numerical calculation we investigate also the
dependence of the LO and NLO QCD corrected total cross sections at
the LHC and the Tevatron on the renormalization and factorization
scales($Q=\mu_r=\mu_f$), and find that the theoretical predictions
including NLO QCD corrections become stable, being nearly
independent of the factorization/renormalization scales for the
processes $pp/p\bar p \to \widetilde{\chi}_1^\pm
\widetilde{\chi}_2^0+X$ as concluded in Ref.\cite{SUSYQCD4}.

\vspace*{1.5cm}
\begin{figure}[hbtp]
\vspace*{-1cm} \centerline{ \epsfxsize = 6.8cm \epsfysize = 7cm
\epsfbox{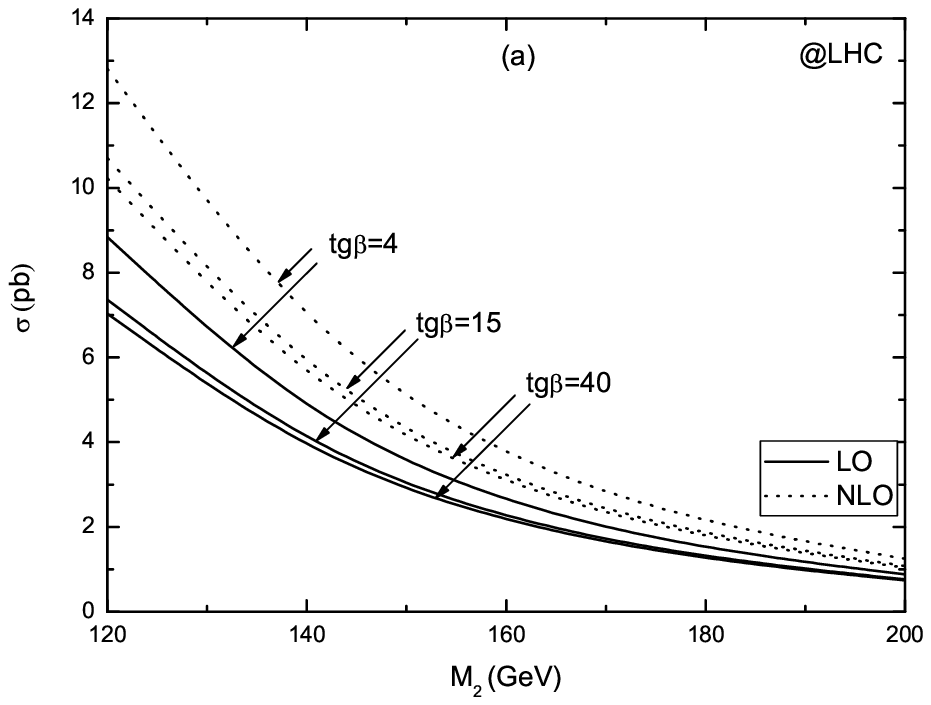} \epsfxsize = 7.5cm \epsfysize = 7cm
\epsfbox{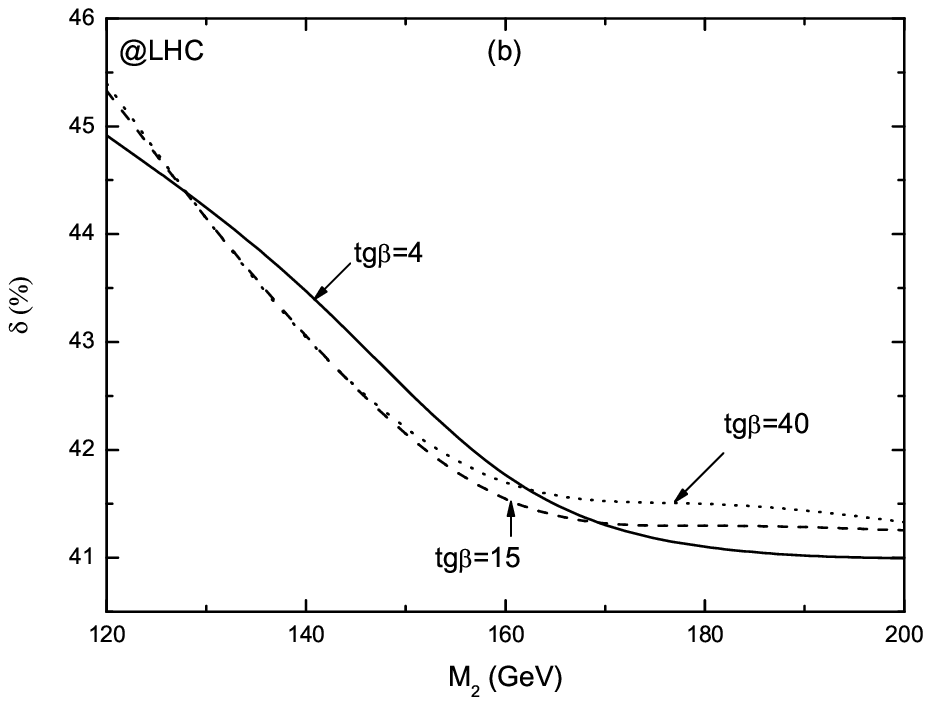}} \vspace*{0.2cm} \centerline{ \epsfxsize =
6.8cm \epsfysize = 7cm \epsfbox{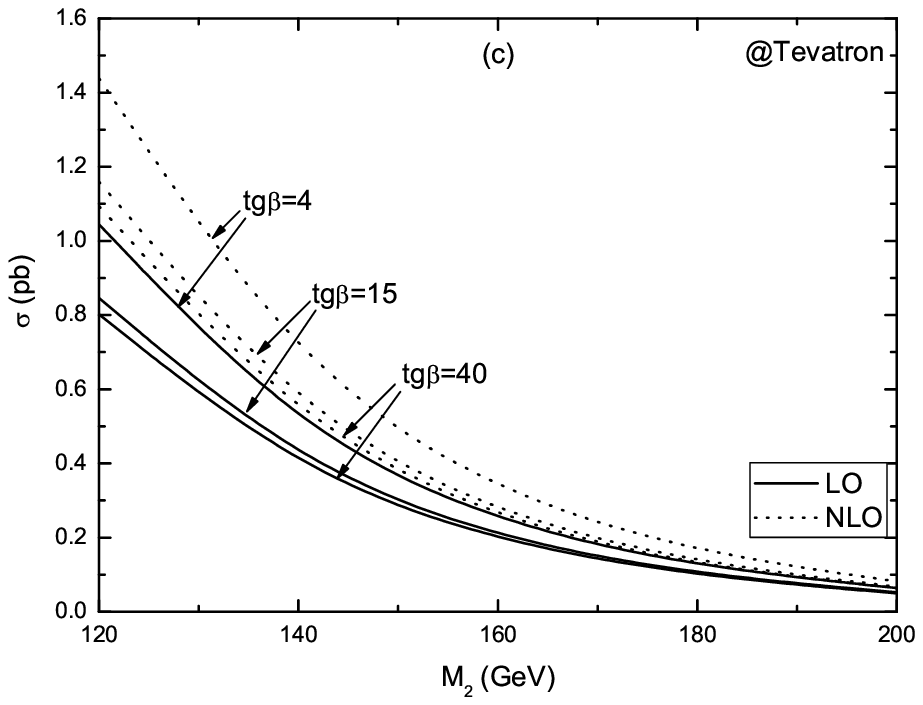} \epsfxsize = 7.5cm
\epsfysize = 7cm \epsfbox{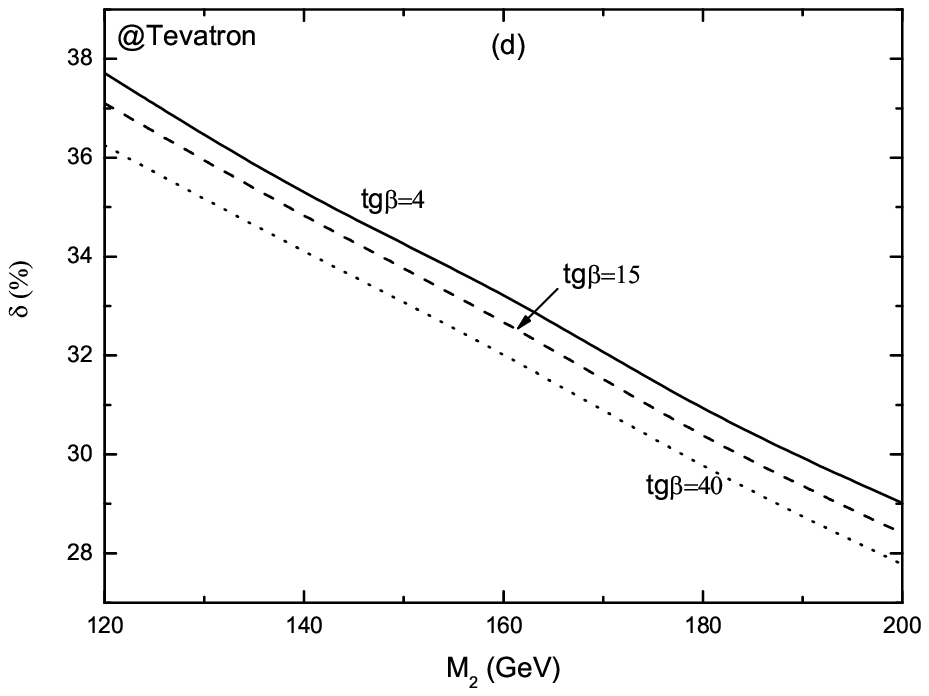}} \vspace*{-0.5cm}\caption{\em
The dependence of the Born, NLO QCD corrected cross sections
(shown in Fig.12(a)) and the corresponding relative corrections
$\delta$ (shown in Fig.12(b)) for the processes $p p/p\bar p \to
\widetilde{\chi}_1^\pm \widetilde{\chi}_2^0+X$ at the LHC and the
Tevatron on the gaugino mass parameter $M_2$. There we take the
input parameters as $m_{A^0}$=300 GeV, $M_{susy}$=500 GeV,
$\mu$=400 GeV and $A_f$=450 GeV, with $\tan\beta=4$,
$\tan\beta=15$ and $\tan\beta=40$
respectively.}{\label{ppudbarxxQCD_m2}}
\end{figure}

\par
In Fig.\ref{ppudbarxxQCD_m2}, the dependence of the Born, NLO QCD
corrected cross sections (shown in Fig.\ref{ppudbarxxQCD_m2}(a)
and (d)) and the relative corrections $\delta$ (shown in
Fig.\ref{ppudbarxxQCD_m2}(b) and (d)) for the processes $pp/p \bar
p \to \widetilde{\chi}_1^\pm \widetilde{\chi}_2^0+X$ at the LHC
and the Tevatron on the gaugino mass parameter $M_2$ are plotted.
There we take the input parameters as $m_{A^0}$=300 GeV,
$M_{susy}$=500 GeV, $\mu$=400 GeV and $A_f$=450 GeV, with
$\tan\beta=4$, $\tan\beta=15$ and $\tan\beta=40$, respectively. We
can see the Born and NLO QCD corrected cross sections decrease
rapidly to a small value with the increment of $M_2$. At the
Tevatron, the NLO QCD relative correction for $\tan\beta=4$
decreases from 37.7$\%$ to 29.0$\%$ with the increment of $M_2$
from $120$ GeV to $200$ GeV. While at the LHC, the relative
corrections decrease at the region $M_2<160$ GeV, and then go down
slowly in the region $M_2>160$ GeV. The NLO QCD relative
corrections for $\tan\beta=4,15,40$ at the LHC have the values in
the range between 45.3$\%$ to 41$\%$, when $M_2$ runs from $120$
GeV to $200$ GeV.

\vspace*{1.5cm}
\begin{figure}[hbtp]
\vspace*{-1cm} \centerline{ \epsfxsize = 6.8cm \epsfysize = 7cm
\epsfbox{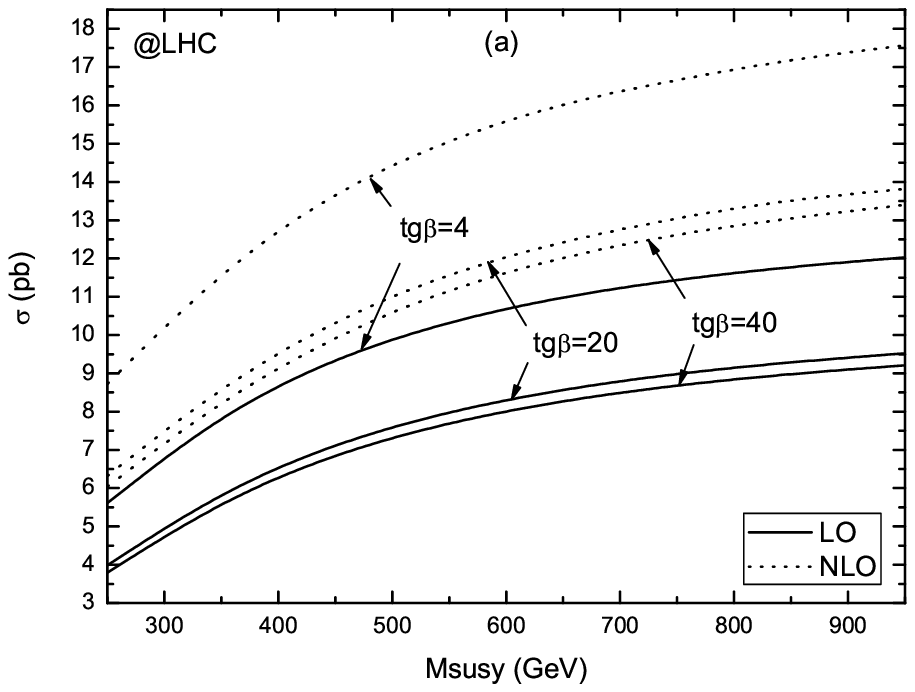} \epsfxsize = 7.5cm \epsfysize = 7cm
\epsfbox{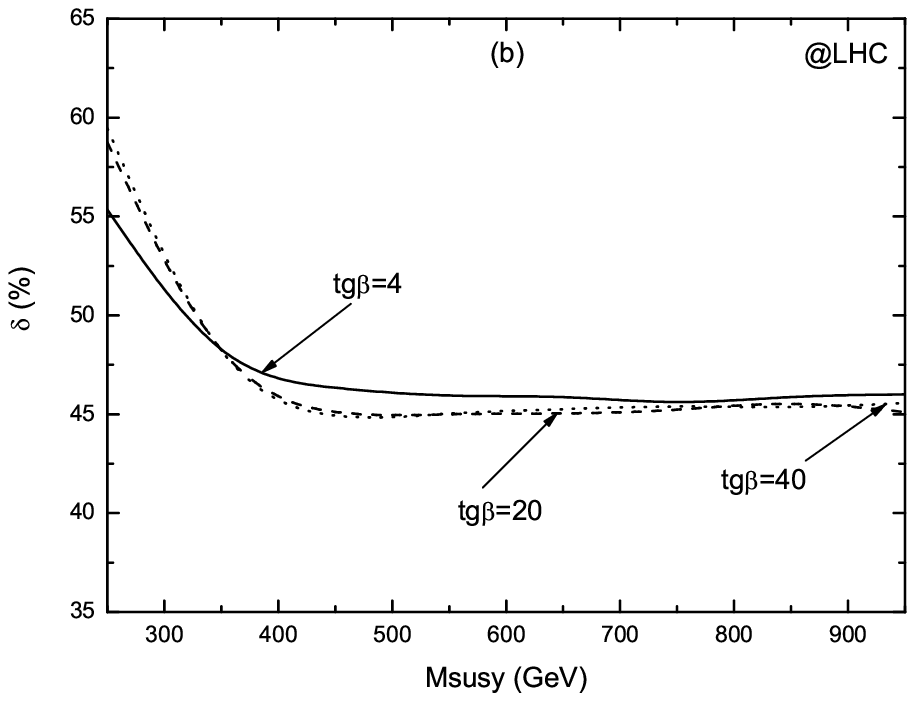}} \vspace*{0.2cm}\centerline{ \epsfxsize =
6.8cm \epsfysize = 7cm \epsfbox{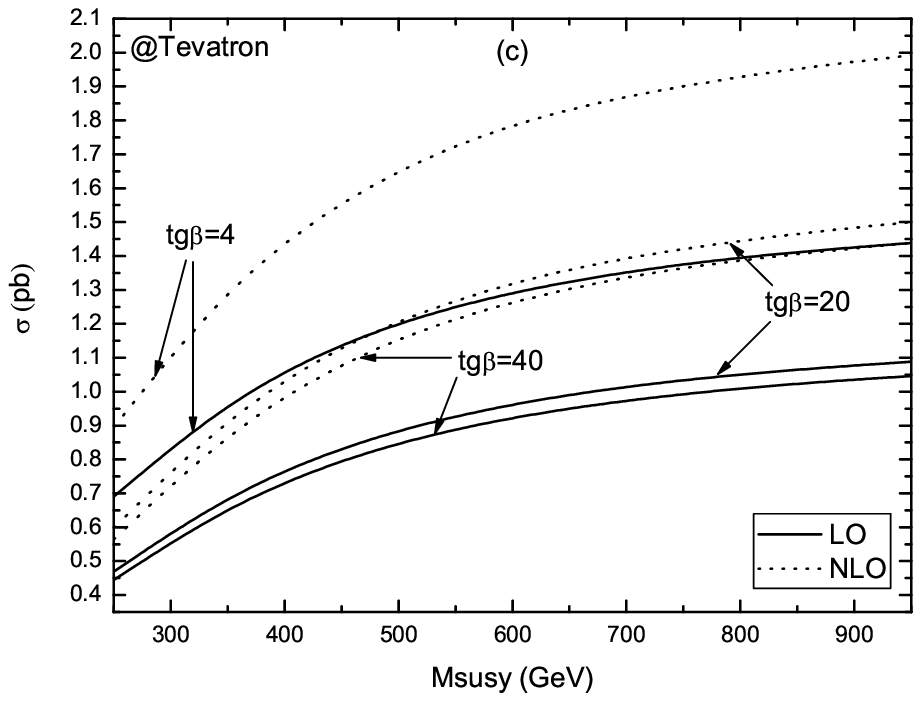} \epsfxsize = 7.5cm
\epsfysize = 7cm \epsfbox{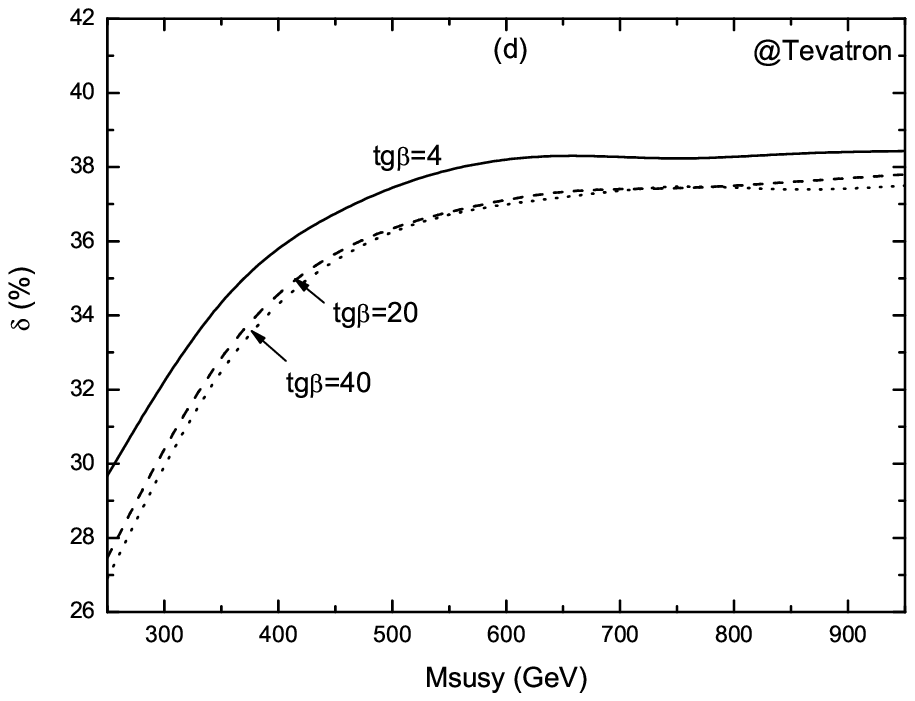}} \vspace*{-0.5cm}\caption{\em
The Born cross sections, NLO QCD corrected cross sections and the
relative NLO QCD corrections $\delta$ for the processes $p \bar
p/pp \to \widetilde{\chi}_1^\pm \widetilde{\chi}_2^0+X$ at the
Tevatron and the LHC as the functions of the supersymmetric mass
parameter $M_{susy}$, on the conditions of $m_{A^0}$=300 GeV,
$M_2$=123 GeV, $\mu$=278 GeV and $A_f$=450 GeV, with
$\tan\beta=4$, $\tan\beta=20$ and $\tan\beta=40$,
respectively.}{\label{ppudbarxxQCD_msusy}}
\end{figure}

\par
In Fig.\ref{ppudbarxxQCD_msusy}(a) and (c), we present the Born
cross sections and NLO QCD corrected cross sections for the
processes $p \bar p/pp \to \widetilde{\chi}_1^\pm
\widetilde{\chi}_2^0+X$ at the Tevatron and the LHC as the
functions of the supersymmetric soft breaking mass parameter
$M_{susy}$, on the conditions of $m_{A^0}$=300 GeV, $M_2$=123 GeV,
$\mu$=278 GeV and $A_f$=450 GeV, with $\tan\beta=5$,
$\tan\beta=20$ and $\tan\beta=40$, respectively. The solid and
dotted curves present the Born and NLO QCD corrected cross
sections, respectively. We can see the cross sections, especially
the corrected ones, increase with the increment of $M_{susy}$.
Fig.\ref{ppudbarxxQCD_msusy}(b) and (d) present the relative
corrections $\delta$ as the functions of $M_{susy}$ at the LHC and
the Tevatron, respectively. The relative correction at the
Tevatron for $\tan\beta=4$ increases rapidly from about 29.7$\%$
to 38.4$\%$ when $M_{susy}$ goes from $250~GeV$ to $600~GeV$, and
becomes to be a constant value about 38.4$\%$ when $M_{susy}$
varies in the region beyond 600 GeV. While at the LHC, the
relative correction decreases rapidly when $M_{susy}$ goes up from
$250~GeV$ to $400~GeV$, and keeps the value about 45$\%$ in the
region of $M_{susy}>400~GeV$.

\par
We also calculate the cross sections of the processes $p p/p\bar p
\to \widetilde{\chi}_1^\pm \widetilde{\chi}_2^0+X$ as the
functions of higgsino-mass parameter $\mu$(with $m_{A^0}=300~GeV$,
$M_2=120~GeV$, $M_{susy}=500~GeV$, $A_f=450~GeV$,
$\tan\beta=4~or~15$ and $\mu\in[250~GeV,~1000~GeV]$), and
$\tan\beta$(with $m_{A^0}=300~GeV$, $M_2=160~GeV$,
$M_{susy}=500~GeV$, $A_f=450~GeV$, $\mu=450~GeV$ and
$\tan\beta\in[5,~40]$) at the LHC/Tevatron and find the result
does not depend much on these parameters. And the relative
corrections have the typical values of about 32$\%$ and 42$\%$ at
the Tevatron and the LHC, respectively.

\par
\section{Full one-loop electroweak corrections to
the \udbarxx subprocess}

\par
For a precise analysis of the association production processes of
neutralino and chargino, higher order electroweak corrections
should be included. The one-loop level UV renormalized electroweak
virtual corrections to the \udbarxx subprocess can be expressed in
the form as
\begin{eqnarray}
\Delta \hat{\sigma}_{vir}(\hat{s},u \bar d \to
\widetilde{\chi}_1^+ \widetilde{\chi}_2^0) =
\frac{1}{8\pi\hat{s}^2}\int^{\hat{t}_{max}}_{\hat{t}_{min}}d\hat{t}Re\overline\sum[({\cal
M}^{V})^{\dagger}{\cal M}^0].
\end{eqnarray}
And again the summation with bar in the equation recalls the same
operations as appeared in Eq.(\ref{summation}). $\hat{t}_{max}$
and $\hat{t}_{min}$ are the same as expressed in
Eq.(\ref{tmaxmin}). ${\cal M}^{V}$ is the the UV renormalized
amplitude of virtual Feynman diagrams including self-energy,
vertex, box and counterterm diagrams. We use ${\it FeynArts}$ and
${\it FormCalc\cite{FAFC}}$ packages to generate Feynman diagrams.
For the numerical evaluation of the loop integrals we use the
developed package ${\it LoopTools}$\cite{LT}.

\subsection{Renormalization scheme}
\par
As we know, the contributions of the electroweak one-loop diagrams
contain both ultraviolet(UV) and infrared(IR) divergences. The UV
divergence can be regularized by adopting the dimensional
reduction($DR$) regularization scheme\cite{DR} and the relevant
fields are renormalized by using the on-shell(OS)
conditions\cite{OMS}\cite{COMS}\cite{ren} (neglecting the finite
widths of particles). In treating the QED soft and collinear
divergences we use again the two cutoff PSS method\cite{PSS}, the
IR divergencies are cancelled in complete analogy to the
calculation for the corresponding QCD radiative corrections as
shown in Section 3. The amplitudes are performed by adopting the
't Hooft-Feynman gauge and $n=4-2\epsilon$ space-time dimensions
to isolate the UV and IR singularities. There is no QED induced
collinear IR singularity from final state radiation due to
chargino being massive, but there exists collinear IR
singularities from the initial state radiation. Similar with that
as declared in Section 3, in the electroweak correction
calculation we use again the fermion flow prescription to deal
with the matrix elements including Majorana
particles\cite{Majorana}. The quark mixing matrix is assumed to be
diagonal. The bare parameters are split into renormalized
parameters and their counter terms.

\par
{\bf 1. Gauge sector}
\par
The definitions and the explicit expressions of the
renormalization constants for the gauge boson sector are written
as\cite{COMS,MSSM}:

\begin{eqnarray}
m^2_{W} \rightarrow  m^2_W + \delta m^2_W, ~~~~~m^2_{Z}
\rightarrow m^2_Z + \delta m^2_Z, \\ W^{\pm} \rightarrow
(1+\frac{1}{2}\delta Z_W)W^{\pm}, ~~~~~~~~~~~~~~\\
\left ( \begin{array}{c}Z
\\ A \end{array}\right ) \rightarrow \left
(\begin{array}{c}1+\frac{1}{2}\delta Z_{ZZ}~~~~~~
\frac{1}{2}\delta Z_{ZZ}\\ \frac{1}{2}\delta
Z_{AZ}~~~~~~1+\frac{1}{2}\delta Z_{AA}
\end{array}\right ) \left (\begin{array}{c}Z \\ A
\end{array}\right ).
\end{eqnarray}

\par
The renormalization conditions are taken as that the renormalized
mass parameters of the physical particles are fixed by the
requirement that they are equal to the physical masses, i.e., to
the real parts of the poles of the corresponding propagators. Then
the relevant fields and mass parameters are properly normalized.
It yields the following results for their counter terms.
\begin{eqnarray}
\delta m^2_Z =\widetilde{Re} \Sigma_T^{ZZ} (m^2_Z),~~~~
\delta m^2_W =\widetilde{Re} \Sigma_T^{WW} (m^2_W),\\
\delta Z_{VV}= -\widetilde{Re}\frac{\partial
\Sigma_T^{VV}(p^2)}{\partial p^2}|_{p^2=0},~~~~~~ V=A, Z, W,\\
\delta Z_{AZ}= -\frac{2 \widetilde{Re}
\Sigma_T^{AZ}(m^2_Z)}{m^2_Z}, ~~~\delta Z_{ZA}= \frac{2
\widetilde{Re} \Sigma_T^{AZ}(0)}{m^2_Z}
\end{eqnarray}
where $\Sigma_T$ denotes the transverse self-energy and
$\widetilde {Re}$ means only taking the real part of the loop
integrals. The weak mixing angle is define as
$s^2_W=1-m^2_W/m^2_Z$\cite{Sirlin}.

\par
{\bf 2. Fermion sector}
\par
The relevant fermion field renormalization constants are defined
as:
\begin{eqnarray}
m_{f,0}=m_f+\delta m_f,~~ f_0^{L}=(1+\frac{1}{2}\delta
Z_{f,ij}^L)f^L, ~~~~ f_0^{R}=(1+\frac{1}{2}\delta Z_{f,ij}^R)f^R.
\end{eqnarray}

For the SM fermions, the normalized constants can be expressed as:
\begin{eqnarray}
\label{eq1} \delta m_f &=& \frac{1}{2} \widetilde{Re}
  \left [ m_f \Sigma^{L}_{f} (m_f^2) + m_f
  \Sigma^{R}_{f}(m_f^2) +
  \Sigma^{S,L}_{f}(m_f^2) +
  \Sigma^{S,R}_{f}(m_f^2)
  \right ],
\end{eqnarray}
\begin{eqnarray}
\delta Z^{L}_{f,ii} &=& - \widetilde{Re}\Sigma^{L}_{f,ii} (m_f^2)
- m_f \frac{\partial}{\partial p^2} \widetilde{Re}
  \left \{ m_f \Sigma^{L}_{f,ii}(p^2)
  + m_f \Sigma^{R}_{f,ii}(p^2) \right. \nonumber\\
&+& \left. \Sigma^{S,L}_{f,ii}(p^2) +
    \Sigma^{S,R}_{f,ii}(p^2) \right \} |_
    {p^2=m_f^2},
\end{eqnarray}
\begin{eqnarray}
\delta Z^{R}_{f,ii} &=& -\widetilde{Re}\Sigma^{R}_{f,ii} (m_f^2) -
  m_f \frac{\partial} {\partial p^2} \widetilde{Re}
  \left\{ m_f \Sigma^{L}_{f,ii}(p^2) +
  m_f \Sigma^{R}_{f,ii}(p^2) \right. \nonumber\\
&+& \left. \Sigma^{S,L}_{f,ii}(p^2) +
    \Sigma^{S,R}_{f,ii}(p^2) \right \} |_
    {p^2=m_f^2},
\end{eqnarray}
\begin{eqnarray}
\delta Z^{L}_{f,ij} &=& \frac{2}{m_{f_i}^2-m_{f_j}^2}
\widetilde{Re} \left [ m_{f_j}^2 \Sigma^{L}_{f,ij} (m_{f_j}^2) +
m_{f_i} m_{f_j}
\Sigma^{R}_{f,ij}(m_{f_j}^2) \right. \nonumber\\
&+&\left.  m_{f_i} \Sigma^{S,L}_{f,ij}(m_{f_j}^2)+ m_{f_j}
\Sigma^{S,R}_{f,ij}(m_{f_j}^2) \right ] ~~~~~~~~~(i\neq j),
\end{eqnarray}
\begin{eqnarray}
\delta Z^{R}_{f,ij} &=& \frac{2}{m_{f_i}^2-m_{f_j}^2}
\widetilde{Re} \left [ m_{f_i} m_{f_j} \Sigma^{L}_{f,ij}
(m_{f_j}^2)+
m_{f_j}^2 \Sigma^{R}_{f,ij}(m_{f_j}^2)  \right. \nonumber\\
&+&\left.   m_{f_j} \Sigma^{S,L}_{f,ij}(m_{f_j}^2)+ m_{f_i}
\Sigma^{S,R}_{f,ij}(m_{f_j}^2) \right ] ~~~~~~~~~(i\neq j).
\end{eqnarray}
The one-particle irreducible two-point function
$i\Gamma_{f,ij}(p^2)$ for fermions is decomposed as
\begin{eqnarray}
\label{eq2}\Gamma_{f,ij}(p^2) &=& \delta_{ij}
         (\rlap/p-m_{f}) +  \left [ \rlap/p P_{L}
    \Sigma_{f,ij}^{L}(p^2)
   + \rlap/p P_{R} \Sigma_{f,ij}^{R}(p^2)\right.   \nonumber\\
   &+& \left. P_{L} \Sigma_{f,ij}^{S,L}(p^2)
    + P_{R} \Sigma_{f,ij}^{S,R}(p^2) \right].
\end{eqnarray}

\par
In our calculation we use an improved scheme to make the
perturbative calculation more reliable. That means we use the
effective $\overline{MS}$ fine structure constant value at $Q=m_Z$
as input parameter,
$\alpha_{ew}(m^2_Z)^{-1}|_{\overline{MS}}=127.918$\cite{alphaEW}.
This results in the counter-term of the electric charge expressed
as \cite{count1, count2, eberl}:
\begin{eqnarray}
\label{ecount}
   \delta Z_e &=&
   \frac{e^2}{6(4\pi)^2}
    \left\{ 4 \sum_f N_C^f e_f^2\left( \Delta+\log\frac{Q^2}{x_f^2} \right)+\sum_{\tilde{f}} \sum_{k=1}^2 N_C^f
    e_{f}^2 \left( \Delta+\log\frac{Q^2}{m^2_{\tilde{f}_k}} \right)     \right.       \nonumber  \\
  &&\left. + 4 \sum_{k=1}^2\left(\Delta+\log\frac{Q^2}{m^2_{\tilde{\chi}_k}}\right)
       +\sum_{k=1}^2\left( \Delta+\log\frac{Q^2}{m^2_{H_k^+}} \right)   \right.      \nonumber   \\
  &&\left.  - 22 \left(\Delta+\log\frac{Q^2}{m_W^2}\right)
  \right\},
\end{eqnarray}
where we take $x_f=m_Z$ when $m_f<m_Z$ and $x_t=m_t$. $e_f$ is the
electric charge of (s)fermion and
$\Delta=2/\epsilon-\gamma+\log4\pi$. $N_C^f$ is color factor,
which equal to 1 and 3 for (s)leptons and (s)quarks, respectively.
In our calculation we take
$Q=(m_{\tilde{\chi}_1^{\pm}}+m_{\tilde{\chi}_1^0})/2$ in using
Eq.(\ref{ecount}).

\par
{\bf 3. Supersymmetric sector}

\par
In the MSSM theory the physical chargino mass eigenstates
$\tilde{\chi}^{\pm}_{1,2}$ are the combinations of charged
gauginos and higgsinos. Their physical masses can be obtained by
diagonalizing the corresponding mass matrix $X$, which has the
form as \cite{UVN}:
\begin{eqnarray}
& & X = \left(
          \begin{array}{cc}
                M_{SU(2)} & \sqrt{2}m_W \sin\beta\\
                \sqrt{2}m_W \cos\beta & \mu
          \end{array}
    \right). \nb
\end{eqnarray}
X is diagonalized with two unitary matrices U and V according to
\begin{eqnarray}
U^*XV^{\dagger}=diag(m_{\tilde\chi_1}, m_{\tilde \chi_2}),
\end{eqnarray}
which yields the chargino masses.

\par
The neutralinos are the mixtures of the neutral gauginos and
higgsinos. Their physical masses can be obtained by diagonalizing
the corresponding mass matrix $Y$ \cite{UVN}.
\begin{eqnarray}
&& Y = \left(
          \begin{array}{cccc}
               M_{U(1)} & 0 & -m_Z s_W \cos\beta & m_Z s_W \sin\beta  \\
               0 & M_{SU(2)} & m_Z c_W \cos\beta & -m_Z c_W \sin\beta \\
               -m_Z s_W \cos\beta & m_Z c_W \cos\beta & 0 & -\mu \\
               m_Z s_W \sin\beta & -m_Z c_W \sin\beta & -\mu & 0
          \end{array}
      \right).
\end{eqnarray}
Y is diagonalized with a unitary matrix N according to
\begin{eqnarray}
N^*YN^\dagger=diag(m_{\tilde\chi^0_1},m_{\tilde\chi^0_2},m_{\tilde\chi^0_3},m_{\tilde\chi^0_4}),
\end{eqnarray}
which yields the four neutralino mass eigenstates.

\par
We follow the renormalization definitions for chargino, neutralino
and sfermion sectors as in Ref.\cite{Eberl}. The chargino,
neutralino wave functions and mass counter terms in the mass
eigenstate basis are introduced as
\begin{equation}
\widetilde{\chi}_i \rightarrow (\delta_{ij}+\frac{1}{2}\delta
\widetilde Z ^{+,0L}_{\tilde{\chi_i}\tilde{\chi_j}} P_L +
\frac{1}{2}\delta \widetilde Z
^{+,0R}_{\tilde{\chi_i}\tilde{\chi_j}} P_R )\widetilde \chi _j,
~~~~~~~m_{\widetilde {\chi} _i} \rightarrow m_{\widetilde {\chi}
_i}+\delta m_{\widetilde {\chi} _i},
\end{equation}
where $\widetilde{\chi}$ stands for both charginos and
neutralinos, $i,j=1,2$ for chargino sector, $i,j=1,2,3,4$ for
neutralino sector. These counter terms can be expressed as the
functions of the corresponding self-energies similar with the
equations of Eqs.(\ref{eq1})-(\ref{eq2}) with the replacements of
$f_{i,j} \rightarrow \tilde{\chi}_{i,j}$.
\par
The wave function and mass counter terms for scalar quarks and
scalar leptons are defined as
\begin{eqnarray} \left ( \begin{array}{c}\tilde{q}_1^{(B)}
\\ \tilde{q}_2^{(B)} \end{array}\right ) \rightarrow \left
(\begin{array}{c}\delta Z^{\tilde{q}1/2}_{11}~~~~~~
\frac{1}{2}\delta Z^{\tilde{q}}_{12}\\ \frac{1}{2}\delta
Z^{\tilde{q}}_{21}~~~~~~\delta Z^{\tilde{q}1/2}_{22}
\end{array}\right ) \left (\begin{array}{c}\tilde{q}_1 \\ \tilde{q}_2
\end{array}\right )= (1+\frac{1}{2}\delta Z^{\tilde{q}})\left (\begin{array}{c}
\tilde{q}_1 \\ \tilde{q}_2 \end{array}\right ),
\end{eqnarray}
\begin{eqnarray} \left ( \begin{array}{c}\tilde{l}_1^{(B)}
\\ \tilde{l}_2^{(B)} \end{array}\right ) \rightarrow \left
(\begin{array}{c}\delta Z^{\tilde{l}1/2}_{11}~~~~~~
\frac{1}{2}\delta Z^{\tilde{l}}_{12}\\ \frac{1}{2}\delta
Z^{\tilde{l}}_{21}~~~~~~\delta Z^{\tilde{l}1/2}_{22}
\end{array}\right ) \left (\begin{array}{c}\tilde{l}_1 \\ \tilde{l}_2
\end{array}\right )= (1+\frac{1}{2}\delta Z^{\tilde{l}})\left (\begin{array}{c}
\tilde{l}_1 \\ \tilde{l}_2 \end{array}\right ),
\end{eqnarray}
where
\begin{eqnarray} \delta Z^{\tilde{q},\tilde{l}} \rightarrow \left
(\begin{array}{c}\delta Z^{\tilde{q},\tilde{l}}_{11}~~~~~~ \delta
Z^{\tilde{q},\tilde{l}}_{12}\\ \delta
Z^{\tilde{q},\tilde{l}}_{21}~~~~~~\delta
Z^{\tilde{q},\tilde{l}}_{22}
\end{array}\right ).
\end{eqnarray}
\par
The corresponding counter terms for the scalar fermions are given
as
\begin{equation}
\delta m^2_{\tilde {q}_i,\tilde{l}_i}=\widetilde {Re} \Sigma
^{\tilde {q}_i,\tilde{l}_i}(m_{\tilde {q}_i,\tilde{l}_i}^2),~~~~
\delta Z_{ii} ^{\tilde {q}_i,\tilde{l}_i}=-\widetilde{Re}
\frac{\partial}{\partial k^2} \Sigma_{ii}^{\tilde {q},\tilde{l}}
(k^2)|_{k^2=m_{\tilde {q},\tilde{l}}^2}.
\end{equation}
\begin{equation}
\delta Z_{ij}^{\tilde {q},\tilde{l}}=-\widetilde{Re} \frac{2
\Sigma_{ij}^{\tilde {q},\tilde{l}} (m_{\tilde
{q}_j,\tilde{l}_j}^2)} {m^2_{\tilde {q}_j,\tilde{l}_j}-m^2_{\tilde
{q}_i,\tilde{l}_i}}~~~~~(i,j=1,2,~~~i \neq j).
\end{equation}
We introduce the counter terms for unitary matrices  $N$, $U$, $V$
and $R$ as follow:
\begin{eqnarray}\label{rc}
N \rightarrow N + \delta N, ~~U \rightarrow U + \delta U, ~~V
\rightarrow V + \delta V,~~ R \rightarrow R + \delta R,
\end{eqnarray}
where $N$ and $R$ are the rotation matrices of neutralino and
squark(slepton) sectors, respectively. $U$ and $V$ are the
diagonal unitary matrices for chargino sector.
\par
The counterterms $\delta U, \delta V$, $\delta N$ and $\delta R$
can be fixed by requiring that the counterterms $\delta U, \delta
V$, $\delta N$ and $\delta R$ cancel the antisymmetric parts of
the wave function
corrections\cite{Eberl}\cite{Denner2}\cite{Kniehl}. We get the
expressions of the counter terms for the neutralino, chargino and
sfermion rotation matrices $N$, $U$, $V$ and $R$ as below:
\begin{eqnarray}
\delta N_{ij}=\frac{1}{4} \sum^4_{k=1}(\delta \widetilde
Z^{0,L}_{ik}-\delta \widetilde Z^{0,R}_{ki})N_{kj},\\ \delta
U_{ij}=\frac{1}{4} \sum^4_{k=1}(\delta \widetilde
Z^{+,R}_{ik}-\delta \widetilde Z^{+,L}_{ki})N_{kj}, ~~~\delta
V_{ij}=\frac{1}{4} \sum^4_{k=1}(\delta \widetilde
Z^{+,L}_{ik}-\delta \widetilde Z^{+,R}_{ki})N_{kj}\\
\delta R^{\widetilde f}_{ij}=\frac{1}{4} \sum^4_{k=1}(\delta
\widetilde Z^ {\widetilde f}_{ik}-\delta \widetilde Z ^{\widetilde
f}_{ki} )R^{\widetilde f}_{kj}
\end{eqnarray}
\par
As we expected, the UV divergence induced by the one-loop diagrams
can be cancelled by that contributed by the counterterm diagrams
exactly. While the soft and collinear IR divergences still exist.

\par
\subsection{Real photon emission}

\par
\begin{figure}[hbtp]
\vspace*{-1cm} \centerline{ \epsfxsize = 8cm \epsfysize = 4cm
\epsfbox{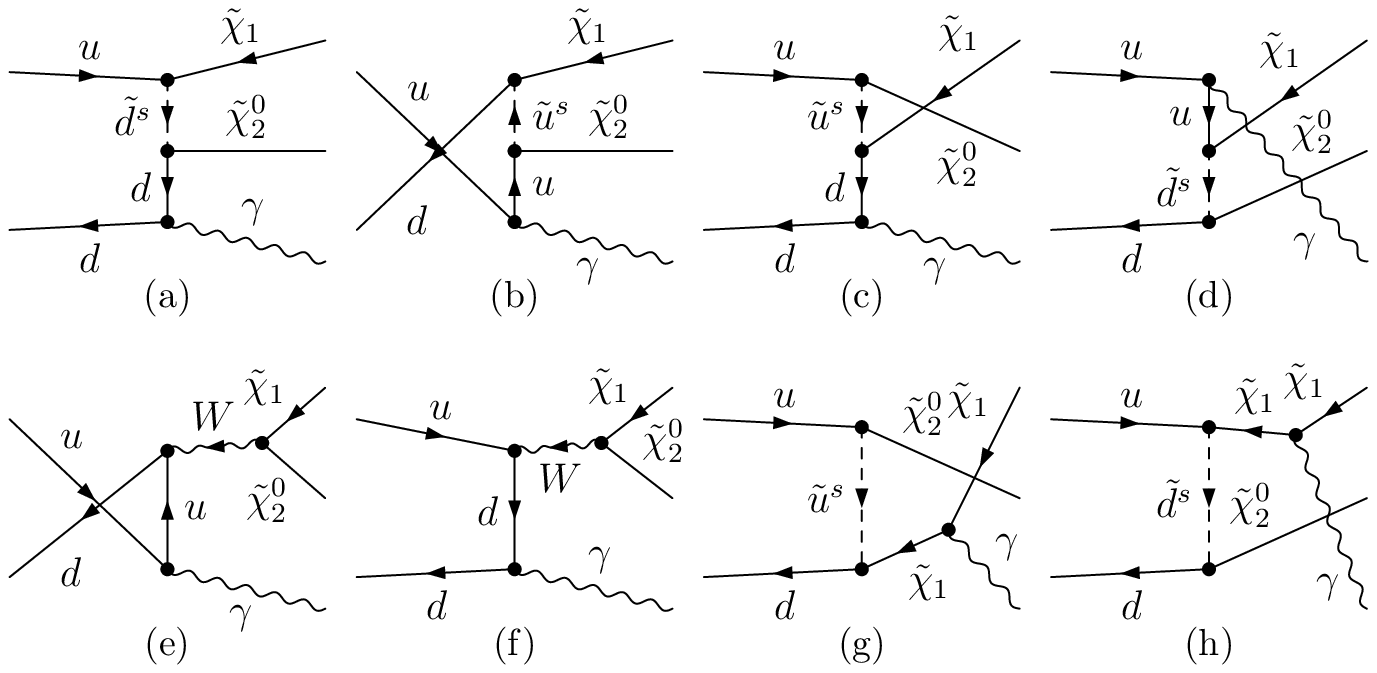}}  \vspace*{0cm}\caption{\em The real photon
emission Feynman diagrams for the subprocess \udbarxxr.}
\label{udbarxxr}
\end{figure}

The soft IR singularity in the ${\cal M}^V$ is originated from
virtual photonic loop correction. It can be cancelled by the
contribution of the real soft photon emission process. The real
photon emission Feynman diagrams for the subprocess \udbarxxr are
shown in Fig.\ref{udbarxxr}.

\vspace*{1.5cm}

\par
We denote the real photon emission process as
\begin{equation}
u(p_1)+\bar d (p_2) \to \widetilde{\chi}_1^+ (k_3) +
\widetilde{\chi}_2^0(k_4) + \gamma(k_5),
\end{equation}
We adopt the general PSS method\cite{PSS} to separate the soft
photon emission singularity from the real photon emission process.
By introducing an arbitrary small soft cutoff $\delta_s$ we
separate the phase space of the subprocess \udbarxxr into two
regions, according to whether the energy of the emitted photon is
soft, i.e. $E_5 \leq \delta_s\sqrt{\hat{s}}/2$, or hard, i.e. $E_5
> \delta_s\sqrt{\hat{s}}/2$, where ${\sqrt{\hat s}}/2$ is the
incoming parton beam energy in the c.m.s. frame. Then the
correction of the real photon emission is broken down into
corresponding soft and hard terms
\begin{equation}
\Delta \hat{\sigma}_{real} =\Delta \hat{\sigma}_{soft}+\Delta
\hat{\sigma}_{hard}
=\hat{\sigma}_0(\hat{\delta}_{soft}+\hat{\delta}_{hard}).
\end{equation}
Although both $\Delta \hat{\sigma}_{soft}$ and $\Delta
\hat{\sigma}_{hard}$ depend on the soft photon cutoff
$\delta_s\sqrt{\hat{s}}/2$, the real correction $\Delta
\hat{\sigma}_{real}$ is cutoff independent. If we use the soft
photon emission approximation\cite{COMS}, we can set $k_5^{\mu}=0$
in the delta function of the phase space element(i.e.,
$\delta^{(n)}(p_1+p_2-k_3-k_4)$) up to corrections of ${\cal
O}(\delta_s)$. Then we can take the n-momenta of the initial and
final particles in the $p_1+p_2$ rest frame as
\begin{eqnarray}\nb
p_1&=&\frac{\sqrt{\hat s}}{2}(1,...,0,0,1),~~ p_2=\frac{\sqrt{\hat
s}}{2}(1,...,0,0,-1),\nb\\
k_3&=&(E_3,...,p\sin\theta,0,p\cos\theta),~~
k_4=(E_4,...,-p\sin\theta,0,-p\cos\theta),\nb\\
k_5&=&E_5(1,...,\sin\theta_1
\sin\theta_2,\sin\theta_1\cos\theta_2,\cos\theta_1).
\end{eqnarray}
Then the integral over the soft photon phase space can be
implemented analytically. We get the expression of the
differential cross section for the subprocess \udbarxx
as\cite{Harris}

\begin{equation}\label{soft cross section}
d\Delta\hat\sigma_{soft} = d\hat \sigma_0 \left
[\frac{\alpha_{ew}}{2\pi}
            \frac{\Gamma(1-\epsilon)}{\Gamma(1-2\epsilon)}
        \left(\frac{4\pi \mu_r^2}{\hat s}\right )^\epsilon \right ]
        \left (\frac{A^S_2}{\epsilon^2}+\frac{A^S_1}{\epsilon}+A^S_0\right ),
\end{equation}
where
\begin{eqnarray}\nb
A^S_2 &=& \frac{5}{9},\\\nb A^S_1 &=& -\frac{4}{9} C_{12}^1+
\frac{12}{9} C_{13}^1+ \frac{6}{9} C_{23}^1 -C_{33}^1,\\\nb A^S_0
&=& -\frac{4}{9} C_{12}^0+ \frac{12}{9} C_{13}^0+ \frac{6}{9}
C_{23}^0-C_{33}^0,
\end{eqnarray}
with
\begin{eqnarray}\nb
C_{12}^1&=&-2\ln{\delta_s},\\\nb
C_{12}^0&=&2 \ln^2{\delta_s}, \nb\\
C_{13}^1&=& -\ln \delta_s -\frac{1}{2} \ln \frac{(E_3-p\cos\theta)^2}{E_3^2-p^2}, \nb \\
C_{13}^0&=& \frac{1}{2}\left [\ln^2
\frac{E_3-p}{E_3-p\cos\theta}-\frac{1}{2}\ln^2
\frac{E_3+p}{E_3-p}+
            2 Li\frac{p\cos\theta-p}{E_3-p}-2 Li \frac{-p\cos\theta-p}{E_3-p\cos\theta}\right ] \nb\\
            &+&\ln \delta_s \ln \frac{(E_3-p\cos\theta)^2}{E_3^2-p^2} +\ln ^2 \delta_s, \nb\\
C_{23}^1&=& -\ln \delta_s -\frac{1}{2} \ln \frac{(E_3+p\cos\theta)^2}{E_3^2-p^2}, \nb \\
C_{23}^0&=& \frac{1}{2}\left [\ln^2
\frac{E_3-p}{E_3+p\cos\theta}-\frac{1}{2}\ln^2
\frac{E_3+p}{E_3-p}+
            2 Li\frac{-p\cos\theta-p}{E_3-p}-2 Li \frac{p\cos\theta-p}{E_3+p\cos\theta}\right ] \nb\\
            &+&\ln \delta_s \ln \frac{(E_3+p\cos\theta)^2}{E_3^2-p^2} +\ln ^2 \delta_s, \nb \\
C_{33}^1&=& -1,\nb\\
C_{33}^0&=& 2\ln \delta_s -\frac{E_3}{p}\ln \frac{E_3+p}{E_3-p},
\end{eqnarray}
In above equations we have following relations between some
variables:
\begin{eqnarray}
E_3&=&\frac{2m_{\tilde{\chi}^+_1}^2-\hat{t}-\hat{u}}{2\sqrt{\hat{s}}},~~~~
p\cos\theta=\frac{\hat{t}-\hat{u}}{2\sqrt{\hat{s}}}, \nb \\
p&=&\frac{1}{2\sqrt{\hat{s}}}\sqrt{(2m_{\tilde{\chi}^+_1}^2+2\sqrt{\hat{s}}m_{\tilde{\chi}^+_1}-\hat
{t}-\hat{u})(2m_{\tilde{\chi}^+_1}^2-2\sqrt{\hat{s}}m_{\tilde{\chi}^+_1}-\hat
{t}-\hat{u})}.
\end{eqnarray}

\par
Since the incoming light-quarks are assumed to be massless in the
parton model and the outgoing particles are massive, there exist
only collinear IR singularities induced by initial state hard
photon radiation. To isolate the collinear singularities from
$\Delta \hat{\sigma}_{hard}$, we further decompose $\Delta
\hat{\sigma}_{hard}$ into a sum of hard collinear(HC) and hard
non-collinear($\overline{\rm HC}$) terms by introducing another
cutoff $\delta_c$ named collinear cutoff
\begin{eqnarray}
\Delta \hat{\sigma}_{hard} =\Delta \hat{\sigma}_{\rm HC}+\Delta
\hat{\sigma}_{\overline{\rm HC}},
\end{eqnarray}
where the HC regions of the phase space are those any one of the
Lorentz invariants $\hat{t}_{15}(\equiv(p_1-k_5)^2)$,
$\hat{t}_{25}(\equiv(p_2-k_5)^2)$ becomes smaller in magnitude
than $\delta_c \hat{s}$ and the emitted photon remains hard.
$\Delta \hat{\sigma}_{\rm HC}$ contains collinear divergences. As
mentioned above, the soft IR divergence of the virtual photonic
corrections can be cancelled exactly by that of soft real
corrections. The remaining collinear singularities are absorbed by
a redefinition(renormalization) of the parton distribution
functions PDFs\cite{Rujula}. This is done in analogy to the
calculation of QCD radiative correction.

\par
In the $\overline{\rm HC}$ region, $\Delta
\hat{\sigma}_{\overline{\rm HC}}$ is finite and can be evaluated
in four-dimensions by applying standard Monte Carlo method. We can
see that $\Delta \hat{\sigma}_{soft}$, $\Delta \hat{\sigma}_{\rm
HC}$ and $\Delta \hat{\sigma}_{\overline{\rm HC}}$, depend on the
two arbitrary parameters $\delta_s$ and $\delta_c$. However, in
the total elecroweak corrected hadronic cross section, after mass
factorization, the dependence on these arbitrary
cutoffs($\delta_s$ and $\delta_c$) cancels, as will be explicitly
shown in numerical calculation(see Section IV.4). This constitutes
an important check of our calculation. Finally, we get an UV and
IR finite corrections $\Delta \sigma$ of the processes $p \bar p/p
p \to \widetilde{\chi}_1^{\pm} \widetilde{\chi}_2^0+X$.

\par
\subsection{The cross sections of processes $p \bar p/p p \to
 \widetilde{\chi}_1^{\pm} \widetilde{\chi}_2^0+X$}

\par
The total electroweak corrected cross sections of the parent
processes $p \bar p/p p \to u \bar d(\bar u d) \to
\widetilde{\chi}_1^{\pm} \widetilde{\chi}_2^0+X$ at the Tevatron
and the LHC can be calculated from the cross sections of
subprocesses \udbarxx and $\bar u d \to \widetilde{\chi}_1^{-}
\widetilde{\chi}_2^0$.
\par
By summing the UV renormalized electroweak virtual corrections and
the real photon emission corrections, the remaining collinear
divergences are absorbed into the redefinition of the distribution
functions. Using the $\overline{\rm MS}$ scheme, the scale
dependent parton distribution functions including ${\cal
O}(\alpha_{ew})$ corrections are given as
\begin{eqnarray}
G_{i/A}(x,\mu_f)=G_{i/A}(x)+\left (-\frac{1}{\epsilon}\right
)\left[\frac{\alpha_{ew}}{2 \pi}
\frac{\Gamma(1-\epsilon)}{\Gamma(1-2 \epsilon)}\left(\frac{4 \pi
\mu_r^2}{\mu_f^2}\right)^{\epsilon}\right]\int^1_z\frac{dz}{z}P_{ij}(z)G_{j/A}(x/z).
\nb \\
\end{eqnarray}
By using above definition, we get a ${\cal O}(\alpha_{ew})$ parton
distribution function counter-terms which are combined with the
hard collinear contributions to result in the $O(\alpha_{ew})$
expression for the remaining collinear contributions:
\begin{eqnarray}
\label{collinear cross section}
 d\sigma^{coll}&=&d\hat{\sigma}^0
\left[\frac{\alpha_{ew}}{2 \pi}
\frac{\Gamma(1-\epsilon)}{\Gamma(1-2 \epsilon)}\left(\frac{4 \pi
\mu_r^2}{\hat{s}}\right)^{\epsilon}\right] \{
\tilde{G}_{u/A}(x_A,\mu_f)G_{\bar{d}/B}(x_B,\mu_f)+G_{u/A}(x_A,\mu_f)\tilde{G}_{\bar{d}/B}(x_B,\mu_f)
\nb \\
&+& \sum_{\alpha=u,\bar{d}}\left [\frac{A_1^{sc}(\alpha \to \alpha
\gamma)}{\epsilon}+A_0^{sc}(\alpha \to
\alpha \gamma)\right ]G_{u/A}(x_A,\mu_f)G_{\bar{d}/B}(x_B,\mu_f) \nb \\
&+& (A\leftrightarrow B)\}dx_Adx_B,
\end{eqnarray}
where $A/B$ are proton/antiproton for the Tevatron, and
proton/proton for the LHC, respectively.
\begin{eqnarray}
A_1^{sc}(u(\bar{d}) \to u(\bar{d}) \gamma)&=& q_f^2(2 \ln
\delta_s+3/2), ~~~A_0^{sc} = A_1^{sc} \ln\left
(\frac{\hat{s}}{\mu_f^2}\right ),
\end{eqnarray}
and
\begin{eqnarray}
\tilde{G}_{\alpha/A,B}(x,\mu_f)=\sum_{c'=\alpha,\gamma}\int^{1-\delta_s
\delta_{\alpha c'}}_x \frac{dy}{y} q_f^2
G_{c'/A,B}(x/y,\mu_f)\tilde{P}_{\alpha c'}(y),~~~(\alpha=u,\bar d)
\end{eqnarray}
with
\begin{eqnarray}
\tilde{P}_{\alpha c'}(y)=P_{\alpha c'} \ln\left
(\delta_c\frac{1-y}{y}\frac{\hat{s}}{\mu_f^2}\right )-P'_{\alpha
c'}(y).
\end{eqnarray}
where $\tilde{P}_{qq}=\frac{3}{4} P_{qq}$ and
$P_{q\gamma}=\frac{1}{3} P_{qg}$, $P_{qq}$ and $P_{qg}$ have the
expressions as shown in Eqs.(\ref{Peq0}) and (\ref{Peq1}),
respectively. $q_f(f=u,\bar d)$ are the charges of quarks.

\par
The final result for the total ${\cal O}(\alpha_{ew})$ correction
consists of two parts of contributions: a two-body term
$\sigma^{(2)}$ and a three-body term $\sigma^{(3)}$. The two-body
correction term $\sigma^{(2)}$ is expressed as
\begin{eqnarray}
\sigma^{(2)}&=&\frac{\alpha_{ew}}{2 \pi} \int
dx_Adx_Bd\hat{\sigma}^0
\{ G_{u/A}(x_A,\mu_f)G_{\bar{d}/B}(x_B,\mu_f)[A^S_0+A^V_0+A_0^{sc}(u\to u\gamma)+A_0^{sc}(\bar d\to \bar d\gamma)] \nb \\
&+& \tilde{G}_{u/A}(x_A,\mu_f)G_{\bar d/B}(x_B,\mu_f)
+G_{u/A}(x_A,\mu_f)\tilde{G}_{\bar d/B}(x_B,\mu_f)+(A
\leftrightarrow B ) \}.
\end{eqnarray}
And the three-body correction term $\sigma^{(3)}$ is written as
\begin{eqnarray}
\sigma^{(3)}&=&\sigma^{(3)}(pp/p\bar p \to u \bar d
\to \widetilde{\chi}_1^+ \widetilde{\chi}_2^0+\gamma)  \nb \\
 &=&\int dx_Adx_B
[G_{u/A}(x_A,\mu_f)G_{\bar{d}/B}(x_B,\mu_f)+(A \leftrightarrow B
)]d\hat{\sigma}^{(3)}( u \bar d
\to \widetilde{\chi}_1^+ \widetilde{\chi}_2^0+\gamma), \nb \\
\end{eqnarray}
where $G$ is the proton/antiproton distribution function. Finally,
the full one-loop electroweak corrected cross section for
$pp/p\bar p \to \widetilde{\chi}_1^+ \widetilde{\chi}_2^0+X$ is
\begin{eqnarray}
\sigma^{EW}=\sigma^{0}+\sigma^{(2)}+\sigma^{(3)}=\sigma^{0}+\Delta
\sigma.
\end{eqnarray}

\par
The cross section part of $\hat\sigma^{(2)}+\hat\sigma^{(3)}$
should be independence of the cutoff parameters $\delta_s$ and
$\delta_c$. The electroweak one-loop relative correction is
defined as $\delta =\Delta \sigma /\hat{\sigma}_0$.

\par
\subsection{Numerical results including electroweak corrections}

\par
In this subsection, we present some numerical results for the
one-loop ${\cal O}(\alpha_{ew})$ electroweak corrections to the
processes $p \bar p/p p \to \widetilde{\chi}_1^{\pm}
\widetilde{\chi}_2^0+X$. We take the SM input parameters as $m_Z =
91.1876~GeV$, $m_W = 80.425~GeV$, $m_t = 178.1~GeV$, $m_b = 4.7~
GeV$\cite{parameter} and neglect the light-quark masses in the
numerical calculation. The fine structure constant is taken having
the value at the $Z^0$-pole, $\alpha_{ew}(m_Z^2)|_{\overline{MS}}
= 1/127.918$\cite{alphaEW}. The new MRST 2004-QED parton
distribution functions including ${\cal O}(\alpha_{ew})$
corrections to the parton evolution are adopted in calculating the
Born and the one-loop order corrected cross sections\cite{Martin}.
The renormalization and factorization scales are taken to be equal
for simplicity($Q=\mu_r=\mu_f$), and have the value being the
average of the final particle masses in analogy to the NLO QCD
calculation. We use again the package FormCalc to obtain all the
masses of supersymmetric particles by inputting the supersymmetric
parameters $\tan\beta$, $m_{A^0}$, $M_{susy}$, $\mu$, $M_2$ and
$A_f$. Among these six input supersymmetric parameters, the CP-odd
Higgs-boson  mass $m_{A^0}$ and $\tan \beta$ with the constraint
$\tan \beta\geq 2.5$ are for the Higgs sector. In FormCalc package
the radiative corrections to Higgs-boson masses up to two-loop
contributions have been involved\cite{Higgsmass}. While the
tree-level Higgs-boson masses can be obtained by using the
equations
\begin{eqnarray}
\label{higgsmass}\nb m^2_{h^0,H^0}&=&\frac{1}{2}\left
(m^2_{A^0}+m^2_{Z^0}\mp
\sqrt{(m^2_{A^0}+m^2_{Z^0})^2-4m^2_{A^0}m^2_{Z^0}\cos^2(2\beta)}\right
),  \\
m^2_{H^\pm}&=&m^2_W+m^2_{A^0}.
\end{eqnarray}
In order to keep the gauge invariance during our numerical
calculation, we adopt the tree-level Higgs-masses obtained from
Eq.(\ref{higgsmass}), but not the Higgs-masses from the output of
FormCalc package through out the tree-level and one-loop
calculations.

\par
\begin{figure}[hbtp] \centerline{
\epsfxsize = 6.8cm \epsfysize = 7cm \epsfbox{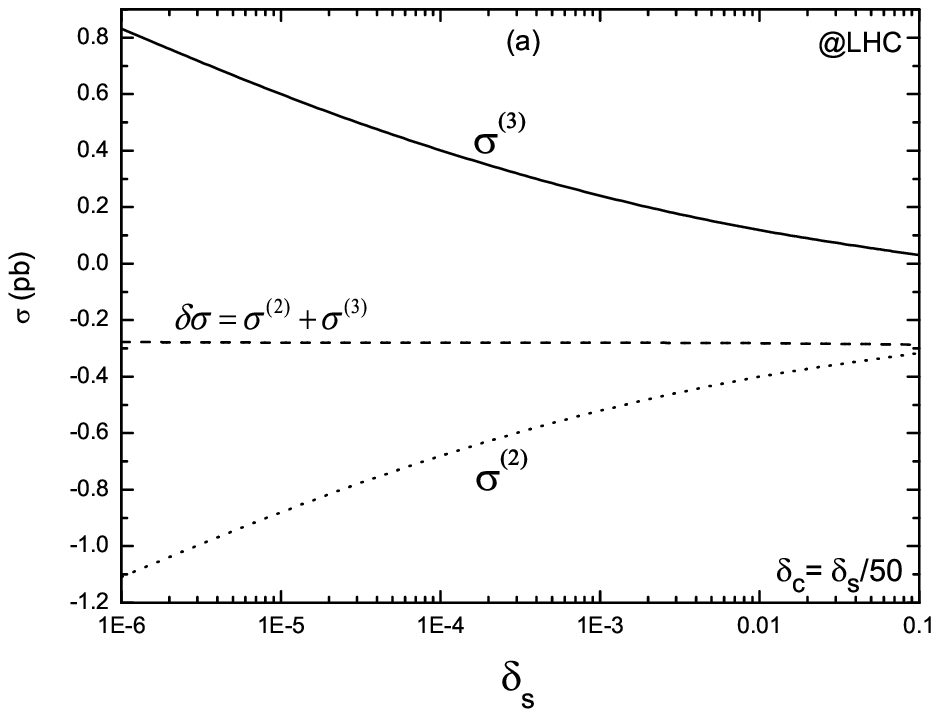}
\epsfxsize = 7.5cm \epsfysize = 7cm
\epsfbox{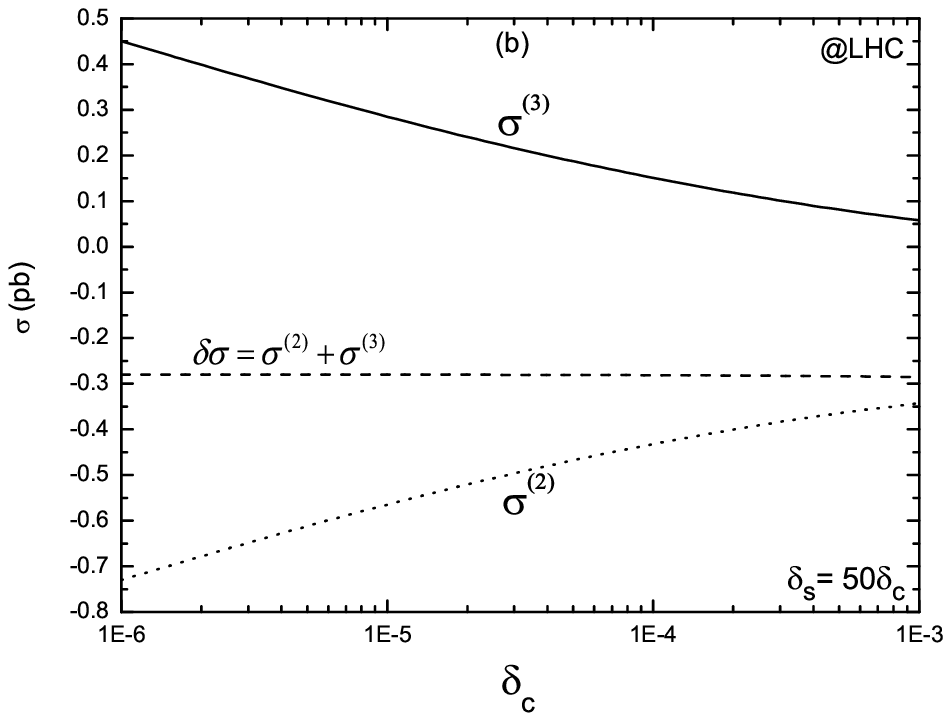}}\vspace*{0.2cm} \centerline{ \epsfxsize =
6.8cm \epsfysize = 7cm \epsfbox{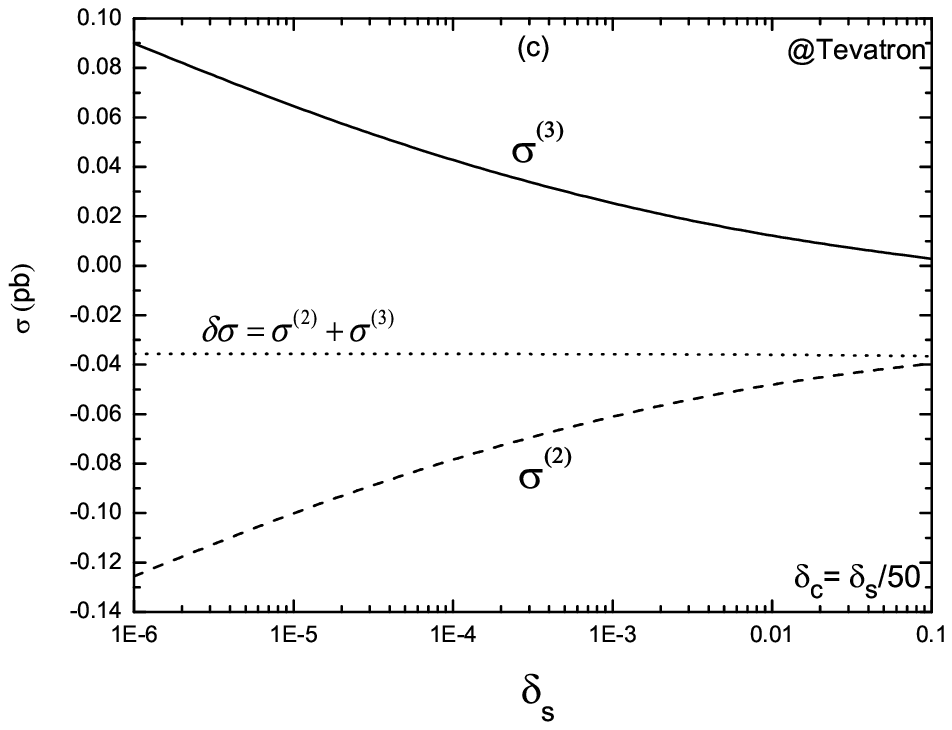} \epsfxsize = 7.5cm
\epsfysize = 7cm \epsfbox{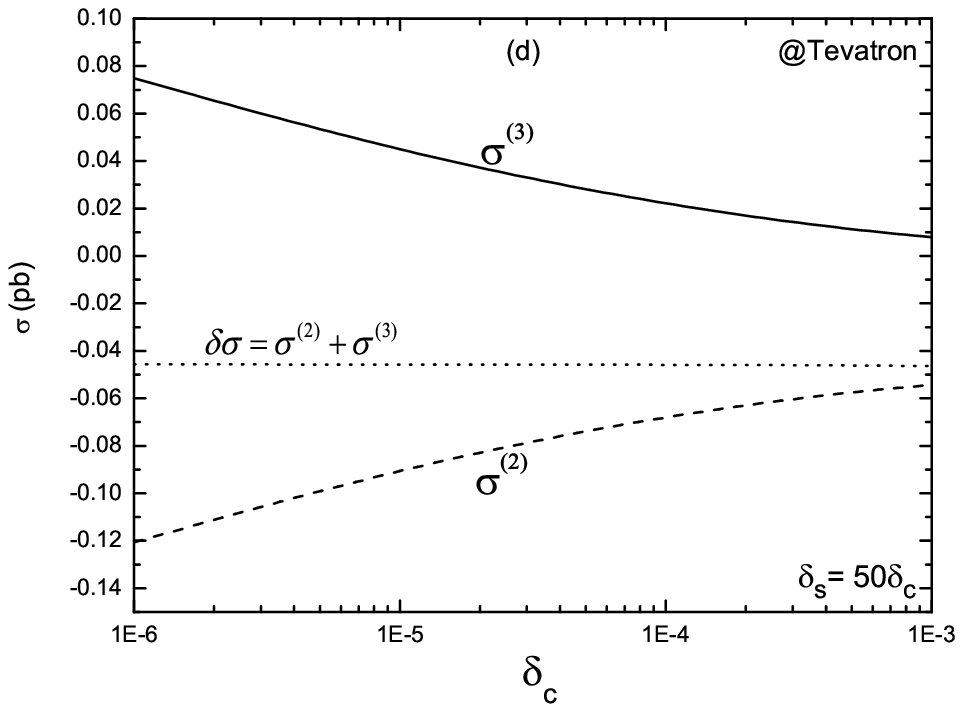}}
\vspace*{-0.5cm}\caption{\em The dependence of the full one-loop
electroweak corrected cross sections for $p \bar p/pp \to
\widetilde{\chi}_1^\pm \widetilde{\chi}_2^0+X$ processes at the
Tevatron and the LHC as the functions of the cutoff $\delta_s$
with $\delta_c=\delta_s/50$(see Fig.\ref{deltasc_ew}(a),(c)) and
$\delta_c$ with $\delta_s=50\delta_c$(see
Fig.\ref{deltasc_ew}(b),(d)), respectively.} \label{deltasc_ew}
\end{figure}

As mentioned above, the final results should be independent on
cutoffs $\delta_s$ and $\delta_c$. For demonstration, we present
the cross section corrections of the $p \bar p/pp \to
\widetilde{\chi}_1^\pm \widetilde{\chi}_2^0+X$ processes as the
functions of the soft cutoffs $\delta_s$ and $\delta_c$ in
Figs.\ref{deltasc_ew}(a-d) on conditions of $\tan\beta=4$,
$m_{A^0}=300$ GeV, $M_{susy}=250$ GeV, $\mu=278$ GeV, $M_2=127$
GeV and $A_f=450$ GeV at the Tevatron and the LHC. The dashed,
solid and dotted lines correspond to the total correction
$\Delta\sigma=\sigma^{(2)}+\sigma^{(3)}$, three-body correction
$\sigma^{(3)}$ and two-body correction $\sigma^{(2)}$,
respectively. As shown in these figures, the full ${\cal O
}(\alpha_{w})$ correction $\Delta\sigma$ is independent of the
soft cutoff $\delta_s$($\delta_c$), as $\delta_s$($\delta_c$)
running from $10^{-6}$($10^{-6}$) to $10^{-1}$($10^{-3}$) and
$\delta_c=\delta_s/50$($\delta_s=50\delta_c$). In the further
numerical calculations, we set $\delta_s=10^{-5}$ and
$\delta_c=\delta_s/50$, if there is no other statement.

\vspace*{1.5cm}

\vspace*{1.5cm}
\begin{figure}[hbtp]
\vspace*{-1cm} \centerline{ \epsfxsize = 6.8cm \epsfysize = 7cm
\epsfbox{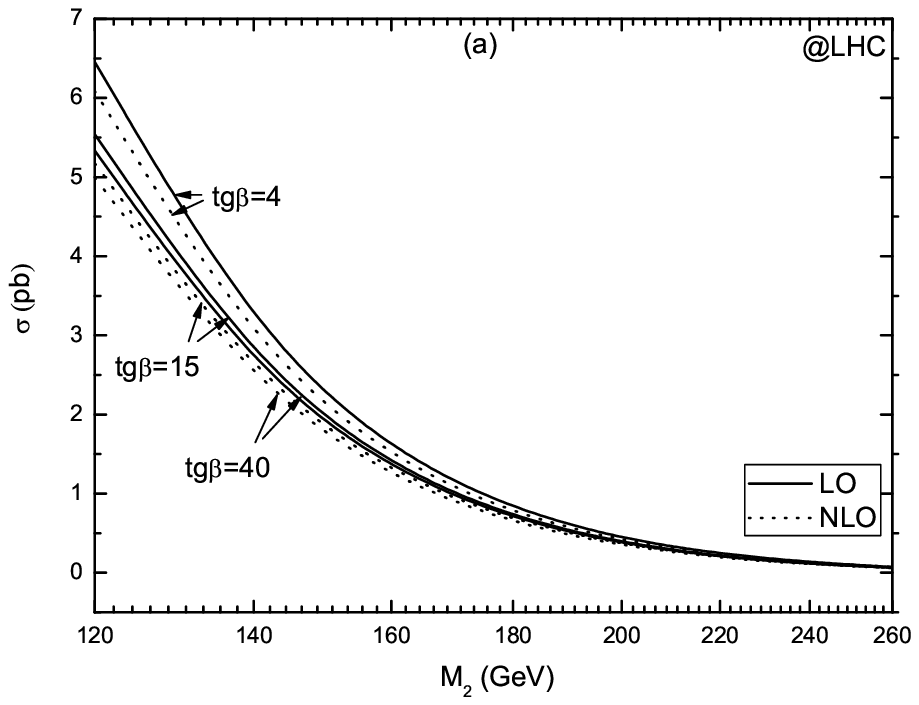} \epsfxsize = 7.5cm \epsfysize = 7cm
\epsfbox{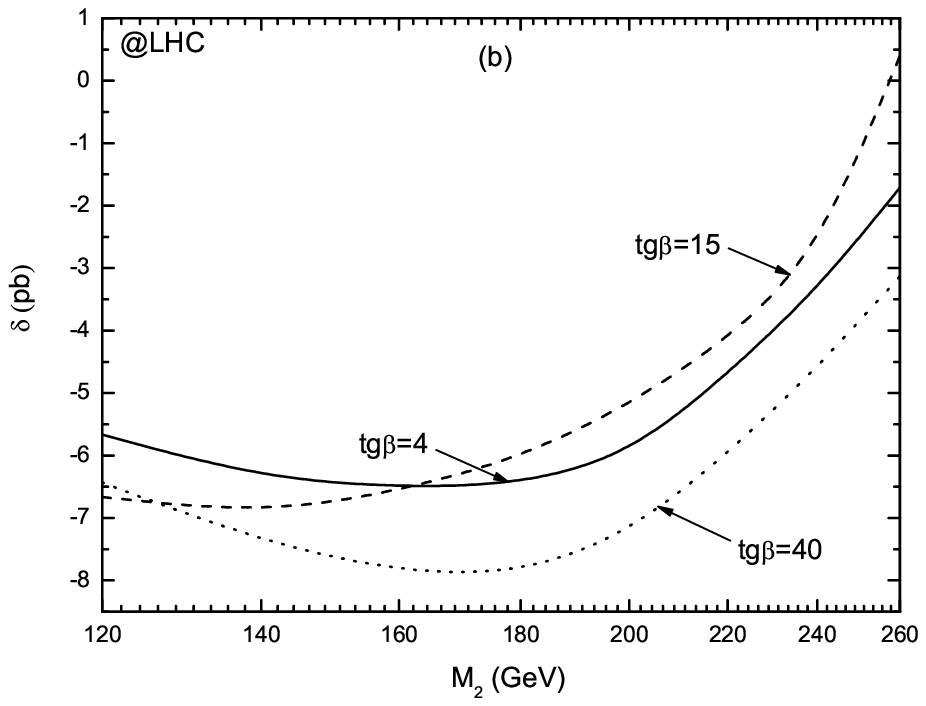}} \vspace*{0.2cm} \centerline{ \epsfxsize =
6.8cm \epsfysize = 7cm \epsfbox{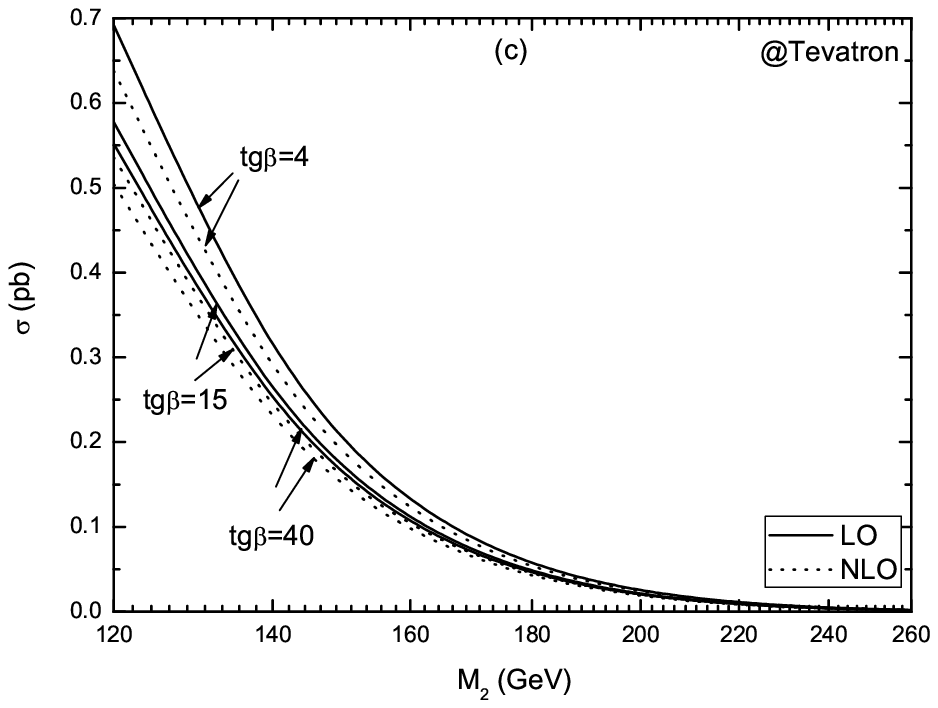} \epsfxsize = 7.5cm
\epsfysize = 7cm \epsfbox{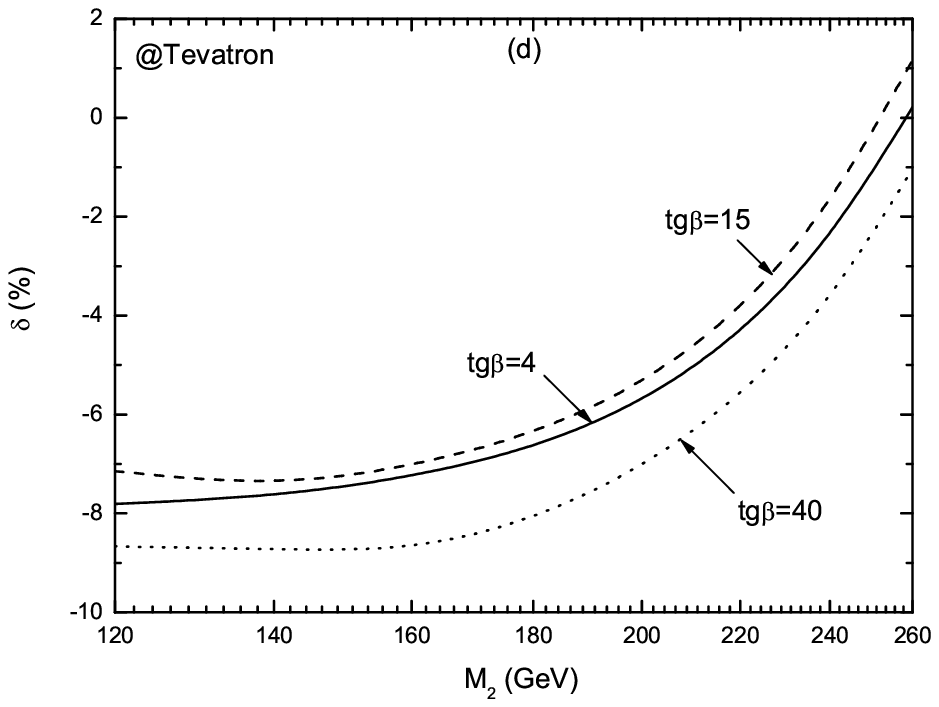}}
\vspace*{-0.5cm}\caption{\em The dependence of the Born, full
one-loop electroweak corrected cross sections (shown in
Fig.\ref{ppudbarxxEW_m2}(a,c)) and the corresponding relative
corrections $\delta$ (shown in Fig.\ref{ppudbarxxEW_m2}(b,d)) for
the processes $p p/p\bar p \to \widetilde{\chi}_1^\pm
\widetilde{\chi}_2^0+X$ at the LHC and the Tevatron on the gaugino
mass parameter $M_2$. There we take the input parameters as
$m_{A^0}$=300 GeV, $M_{susy}$=500 GeV, $\mu$=400 GeV and $A_f$=450
GeV, with $\tan\beta=4$, $\tan\beta=15$ and $\tan\beta=40$
respectively.}\label{ppudbarxxEW_m2}
\end{figure}

\par
In Fig.\ref{ppudbarxxEW_m2}, the dependence of the Born cross
sections, full one-loop \EW corrected cross sections and the
corresponding relative corrections of processes $pp/p\bar p \to
\widetilde{\chi}_1^\pm \widetilde{\chi}_2^0+X$ at the LHC and the
Tevatron on the gaugino mass parameter $M_2$ are depicted. There
we take the input parameters as $m_{A^0}$=300 GeV, $M_{susy}$=350
GeV, $\mu$=550 GeV and $A_f$=450 GeV, with $\tan\beta=4$,
$\tan\beta=15$ and $\tan\beta=40$, respectively. We can see that
the Born and electroweak corrected cross sections decrease rapidly
to small values with $M_2$ going from 120 GeV to 260 GeV at the
Tevatron (shown in Fig.\ref{ppudbarxxEW_m2}(c)) and the LHC(shown
in Fig.\ref{ppudbarxxEW_m2}(a)) due to kinematical effects,
because the masses of the final-state chargino and neutralino are
roughly proportional to $M_2$. While the relative correction
$\delta$(shown in Fig.\ref{ppudbarxxEW_m2}(b)(d)) increases
clearly. At the Tevatron the relative correction $\delta$ is in
the range from -8.8$\%$ to 1.2$\%$ with our chosen parameters,
while at the LHC it can reach -7.9$\%$. Here we have the masses of
chargino and neutralino in the ranges of $m_{\tilde \chi ^+_1}\in
[111.756~GeV, 252.504~GeV]$, $m_{\tilde \chi ^0_2}\in
[112.067~GeV,252.538~GeV]$.

\vspace*{1.5cm}
\begin{figure}[hbtp]
\vspace*{-1cm} \centerline{ \epsfxsize = 6.8cm \epsfysize = 7cm
\epsfbox{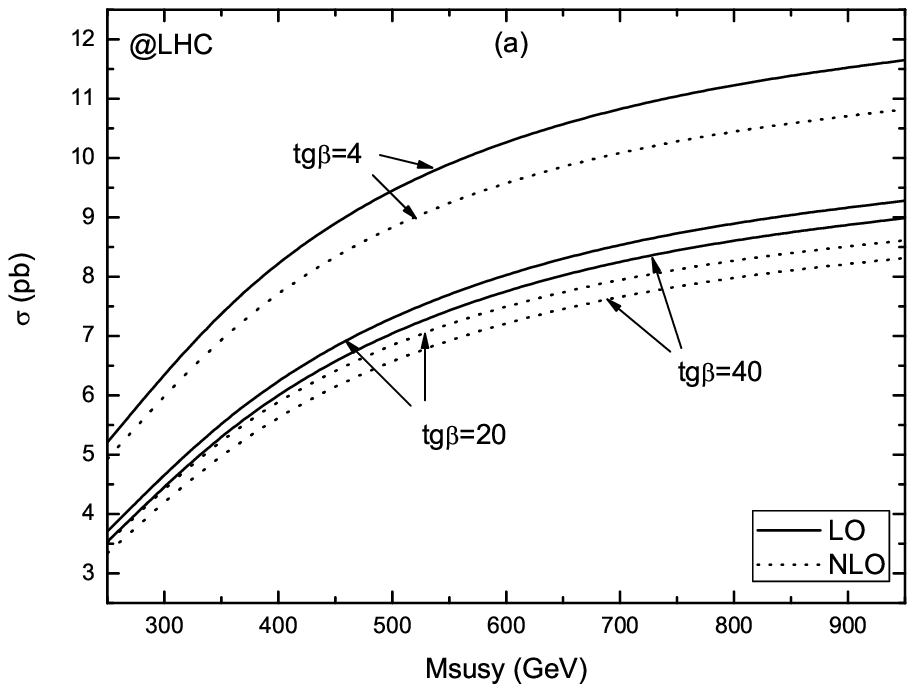} \epsfxsize = 7.5cm \epsfysize = 7cm
\epsfbox{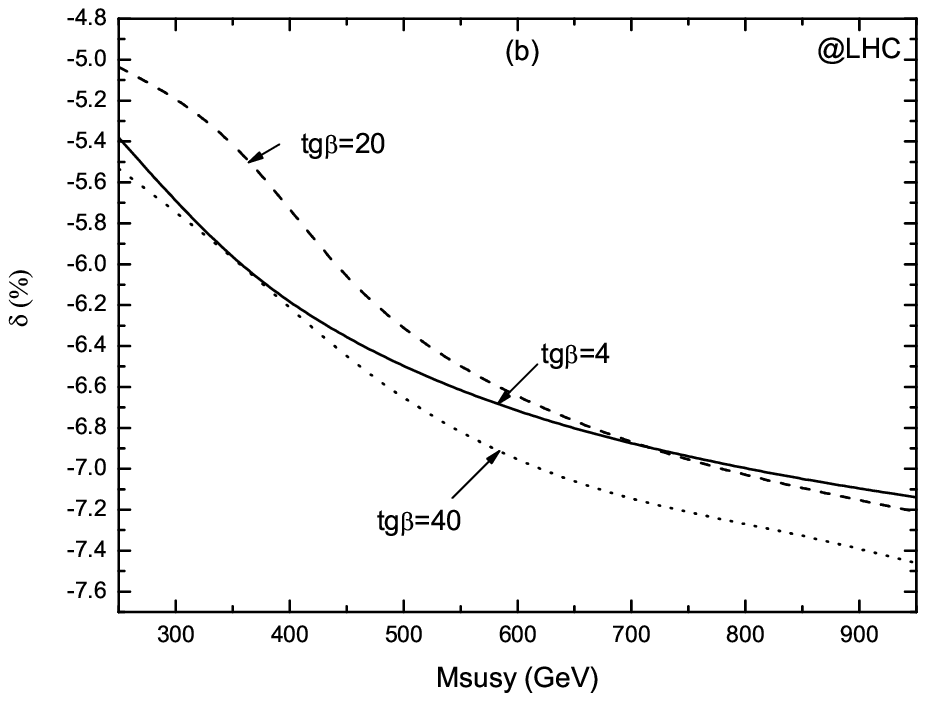}}\vspace*{0.2cm} \centerline{ \epsfxsize =
6.8cm \epsfysize = 7cm \epsfbox{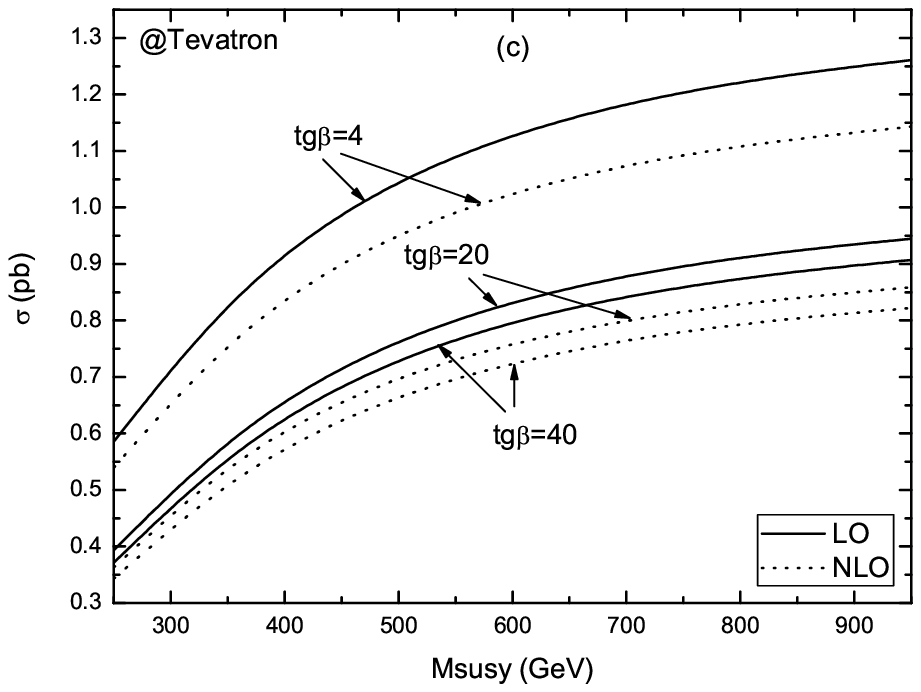} \epsfxsize = 7.5cm
\epsfysize = 7cm \epsfbox{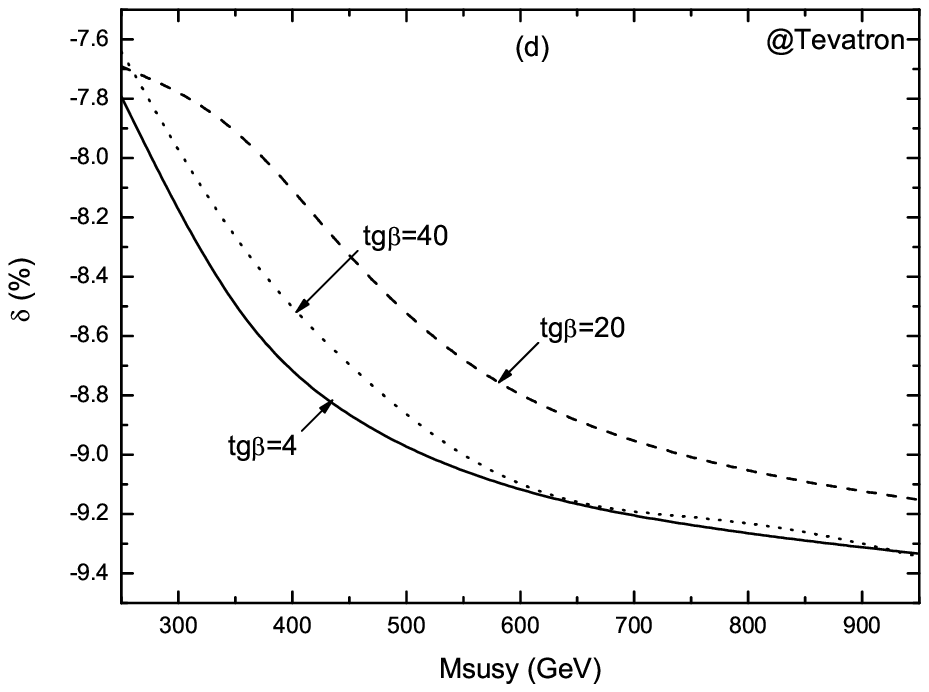}}
\vspace*{-0.5cm}\caption{\em The dependence of the Born and the
electroweak corrected cross sections of process $pp/p\bar p \to
\widetilde{\chi}_1^\pm \widetilde{\chi}_2^0+X$(see in
Fig.\ref{ppudbarxxEW_msusy}(a)(c)) and the corresponding relative
corrections{see in Fig.\ref{ppudbarxxEW_msusy}(b)(d)} on the SUSY
soft breaking mass parameter $M_{susy}$ at the LHC and the
Tevatron on the conditions of $m_{A^0}$=300GeV, $M_2$=127 GeV,
$\mu$=278 GeV and $A_f$= 450GeV, with $\tan\beta=4$,
$\tan\beta=20$ and $\tan\beta=40$,
respectively.}{\label{ppudbarxxEW_msusy}}
\end{figure}

\par
In Figs.\ref{ppudbarxxEW_msusy}(a,c), we show the dependence of
the Born and the full one-loop electroweak corrected cross
sections of processes $pp/p\bar p \to \widetilde{\chi}_1^\pm
\widetilde{\chi}_2^0+X$ on the SUSY soft breaking mass parameter
$M_{susy}$ at the LHC and the Tevatron, on the conditions of
$m_{A^0}$=300 GeV, $M_2$=127 GeV, $\mu$=278 GeV and $A_f$=450 GeV,
with $\tan\beta=4$($m_{\tilde \chi^+_1}$=104.428 GeV, $m_{\tilde
\chi^0_2}$=106.632 GeV), $\tan\beta=20$($m_{\tilde
\chi^+_1}$=113.353 GeV, $m_{\tilde \chi^0_2}$=113.780 GeV) and
$\tan\beta=40$($m_{\tilde \chi ^+_1}$=114.577 GeV, $m_{\tilde \chi
^0_2}$=114.839 GeV), respectively. The solid curves represent the
Born cross sections and the dotted curves represent the one-loop
electroweak corrected cross sections. We can see the total cross
sections have the values from $0.34~pb$ to $1.26~pb$ and from
$3.33~pb$ to $11.65~pb$ at the Tevatron and the LHC respectively,
as the SUSY soft breaking mass parameter $M_{susy}$ runs from 250
GeV to 950 GeV. Figs.\ref{ppudbarxxEW_msusy}(b,d) present the
one-loop electroweak relative corrections as the functions of
$M_{susy}$, and we can see the relative corrections generally
decrease with the increment of $M_{susy}$. And the relative
corrections are between $-9.33\%$ and $-7.64\%$ at the Tevatron
and between $-7.46\%$ and $-5.03\%$ at the LHC with our chosen
parameters.

\par
We also calculate the Born cross sections, the one-loop
electroweak corrected cross sections and the corresponding
relative corrections as the functions of $\mu$ (with
$m_{A^0}=300~GeV$, $M_2=127~GeV$, $M_{susy}=250~GeV$, and
$A_f=450~GeV$, $\tan\beta=4~or~15$ and
$\mu\in[250~GeV,1000~GeV]$), and $\tan\beta$ (with
$m_{A^0}=300~GeV$, $M_2=200~GeV$, $M_{susy}=350~GeV$,
$A_f=450~GeV$, $\mu=550~GeV$ and $\tan\beta\in[5,40]$). We find
that the results do not depend much on those parameters. The
relative corrections have the typical values of about $-8\%$ and
$-6\%$ at the Tevatron and the LHC, respectively.

\par
\section{Summary}

\par
In this paper, we present the calculations of the NLO QCD and the
full one-loop electroweak corrections to the processes $p \bar
p/pp \to \widetilde{\chi}_1^{\pm} \widetilde{\chi}_2^0+X$ at the
Tevatron and the LHC in the framework of the MSSM. In the
calculations of both the NLO QCD and one-loop electroweak
corrections we apply the algorithm of the phase-space slicing(PSS)
method. We analyze the numerical results and investigate the
dependence of the cross sections and corresponding relative
corrections for the processes on several supersymmetric
parameters. We find that the NLO QCD corrections generally
increase the Born cross section, while the electroweak corrections
decrease the Born cross sections in most of the chosen parameter
space. The contributions from the NLO QCD corrections make the
theoretical predictions nearly independent of the renormalization
and factorization scales. Our results show that the NLO QCD and
electroweak relative corrections typically have the values of
about 32$\%$(42$\%$) and -8$\%$(-6$\%$) at the Tevatron(LHC),
respectively. We conclude that the NLO QCD and complete one-loop
electroweak corrections to the processes $p \bar p/pp \to
\widetilde{\chi}_1^{\pm} \widetilde{\chi}_2^0+X$ are generally
significant and should be considered in high precision analysis.

\vskip 5mm
\par
\noindent{\large\bf Acknowledgments:}

This work was supported in part by the National Natural Science
Foundation of China and a special fund sponsored by Chinese
Academy of Science.

\end{document}